\documentclass[aps,prl,amsmath,amssymb,reprint,superscriptaddress]{revtex4-2}

\usepackage{hyperref}
\hypersetup{
	colorlinks,
	linkcolor={rgb:red,2; blue,1},
	citecolor={blue!60!black},
	urlcolor={blue!60!black}
}
\usepackage{graphicx}
\usepackage{amsmath,amssymb,amsthm,bbm,bm}
\usepackage{xcolor}
\usepackage{txfonts}
\usepackage{braket}  

\usepackage{bbold}

\renewcommand{\dag}{\dagger}

\renewcommand{\H}{\hat{H}}

\newcommand{\ii}{\ensuremath{\mathrm{i}}}
\newcommand{\ee}{\ensuremath{\mathrm{e}}}
\newcommand{\dd}{\mathrm{d}}

\newcommand{\Tr}{\mathrm{Tr}}

\renewcommand{\a}{\ensuremath{\hat{a}}}
\newcommand{\adag}{\ensuremath{\hat{a}^\dagger}}
\renewcommand{\b}{\ensuremath{\hat{b}}}
\newcommand{\bdag}{\ensuremath{\hat{b}^\dagger}}
\renewcommand{\c}{\ensuremath{\hat{c}}}
\newcommand{\cdag}{\ensuremath{\hat{c}^\dagger}}

\renewcommand{\H}{\ensuremath{\hat{H}}}

\graphicspath{{./figs/}}

\begin{document}
	
	\title{Robust nonequilibrium edge currents with and without band topology}
	\author{Mark T. Mitchison}
	\email{mark.mitchison@tcd.ie}
	\affiliation{School of Physics, Trinity College Dublin, College Green, Dublin 2, Ireland}
	\author{\'Angel Rivas }
	\email{anrivas@ucm.es}
	\author{Miguel A. Martin-Delgado}
	\email{mardel@ucm.es}
	\affiliation{Departamento de F\'{\i}sica Te\'orica, Facultad de Ciencias F\'isicas, Universidad Complutense, 28040 Madrid, Spain.}
	\affiliation{CCS-Center for Computational Simulation, Campus de Montegancedo UPM, 28660 Boadilla del Monte, Madrid, Spain.}
	
	\begin{abstract}
		We study two-dimensional bosonic and fermionic lattice systems under nonequilibrium conditions corresponding to a sharp gradient of temperature imposed by two thermal baths. In particular, we consider a lattice model with broken time-reversal symmetry that exhibits both topologically trivial and nontrivial phases. Using a nonperturbative Green function approach, we characterize the nonequilibrium current distribution in different parameter regimes. For both bosonic and fermionic systems, we find chiral edge currents that are robust against coupling to reservoirs and to the presence of defects on the boundary or in the bulk. This robustness not only originates from topological effects at zero temperature but, remarkably, also persists as a result of dissipative symmetries in regimes where band topology plays no role. Chirality of the edge currents implies that energy locally flows against the temperature gradient without any external work input. In the fermionic case, there is also a regime with topologically protected boundary currents, which nonetheless do not circulate  around all system edges. 
		
	\end{abstract}
	\maketitle
	
	The physics of boundary structures, whether they be dots, lines or surfaces, has attracted a great deal of attention from various directions in the past.  The theoretical discovery, and its subsequent experimental verification, of both insulating and superconducting topological materials \cite{Haldane,KaneMele,Bernevig,ExpZhang,QiZhang,Hasan} has further spurred the study of boundary physics. The main reason for this is that nontrivial band topology endows edge phenomena with a remarkable robustness. This feature opens up a myriad of possible applications that go well beyond condensed matter physics.  Historically, most of the mainstream studies conducted on those topological materials shared two basic properties in common:
	(i)~the quantum system is considered as closed and thus isolated from the detrimental effects of the surrounding environment; (ii)~the constituent particles are fermions. The combination of these two properties underpins the stability of boundary effects in topological phases of matter.
		
	In this work, we present a different paradigm of robust boundary physics in which we depart from these two common features, entering less well-trodden ground. The motivation is to study more demanding scenarios corresponding to a quantum system coupled to thermal baths, which generate external noise yielding fluctuations and dissipation \cite{ Breuer2002, GardinerZoller, Kamenev2009,RivasHuelga,Landi2021}. In fact, understanding noisy circumstances
	like these is crucial for the successful development of scalable quantum technologies. In this context, we find that robust edge currents can be generated without resorting to the standard band topological mechanism, focusing our attention not only on fermionic systems, but on bosonic particles as well.
	
	To illustrate this unusual form of dissipative robustness, we consider a bosonic variant of the model previously introduced by Qi, Wu and Zhang (QWZ) for fermions on a square two-dimensional (2D) lattice~\cite{Qi2006}. This model has the virtue of presenting two topologically different band structures depending on the values of its coupling parameters. In the fermionic case at half filling, these two correspond to different phases: one is a trivial insulator, and the other a topological insulator. Of course, the single-particle band structure of the QWZ model is the same for fermions and bosons. However, the statistics of the particles determine how those bands are filled.  The Pauli principle forces fermions to fill the bands up to the Fermi level, thereby unveiling the band topology. On the contrary, bosons at low temperatures tend to condense in the single-particle ground state, making them largely insensitive to the global band structure.

	\begin{figure}[b]
		\includegraphics[width=\linewidth]{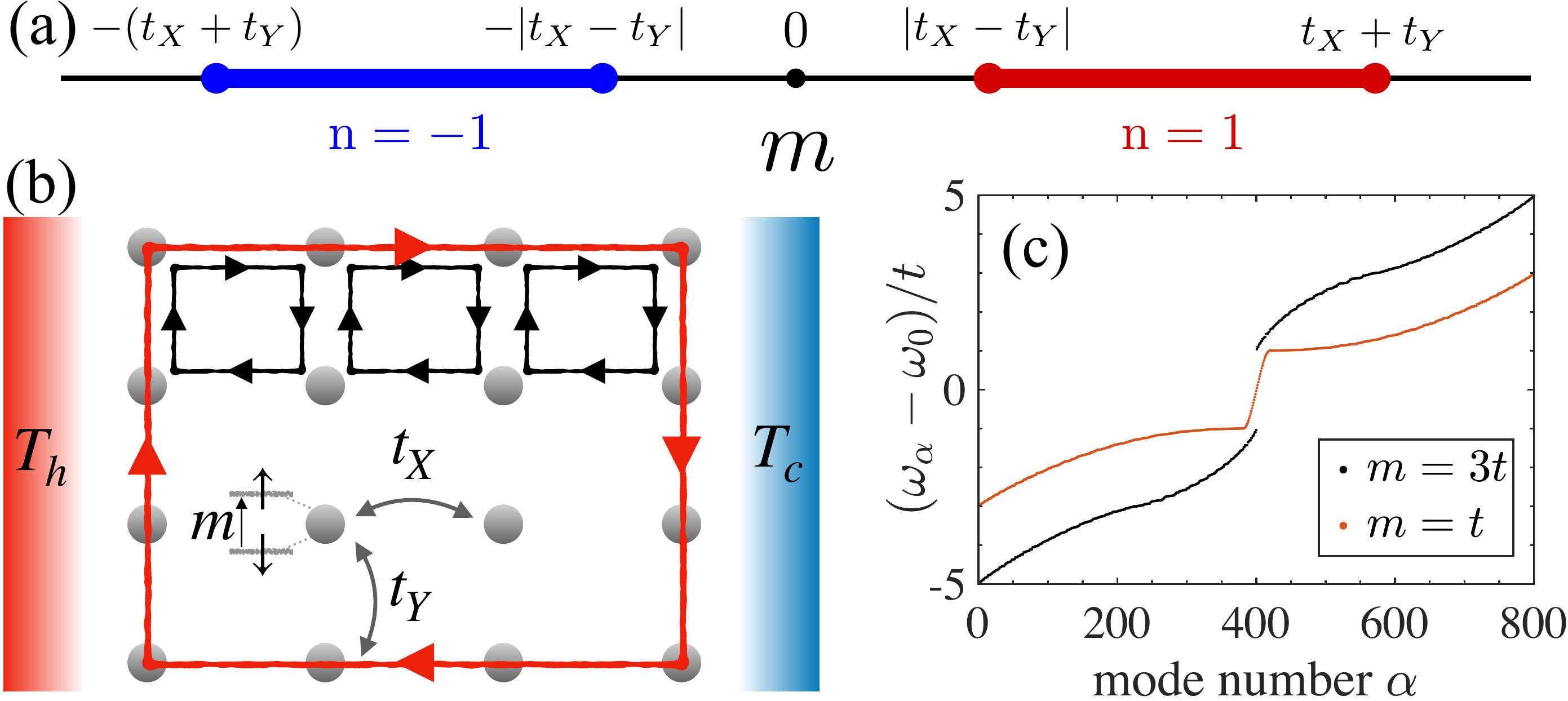}
		\caption{The QWZ model. (a) Phase diagram showing the topologically nontrivial regimes with nonzero Chern number $\mathrm{n}=\pm 1$, a topological invariant of the band structure under periodic boundary conditions~\cite{Hasan}. (b)~Schematic of the system, featuring a 2D lattice hosting particles with two internal states, coupled to thermal baths at different temperatures. An edge current (red) emerges from an erasure effect where modes with the same current circulation (black) cancel in the bulk but add constructively on the boundary. (c) Single-particle eigenenergies $\omega_\alpha$ as a function of the mode index $\alpha$ for a system with open boundary conditions, with $L_X=L_Y=20$, $t_X=t_Y=t$.
			\label{fig:QWZ}}
	\end{figure}
	
	Remarkably, for both bosonic and fermionic QWZ lattices, we find dissipatively robust, chiral edge currents flowing between two thermal reservoirs in a parameter regime leading to trivial band topology. Here, robustness is defined by stability of the currents with respect to the introduction of defects on the edges or in the bulk. This stability is not arbitrary, but is subject to the fulfilment of discrete symmetries on the geometrical distribution of the defects, which must be compatible with the underlying symmetries of the non-equilibrium steady state (NESS). In the fermionic case, and when the temperature of one of the two reservoirs is much smaller than the band gap, we also observe topologically protected edge currents that are effectively independent of the defect distribution but may not circulate around every edge of the perturbed system.
	
The interplay between topology and symmetries in dissipative quantum systems has recently been explored within a Markovian approximation~\cite{Diehl2011,Viyuela2012,Rivas2013,Budich2015,Iemini2016,Linzner2016,Rivas2017,Kawabata2019,Song2019,Shavit2020,Gau2020,Lieu2020,McGinley2020,Flynn2021}, which generally requires weak system-reservoir interactions (see also Ref.~\cite{Altland2021} for non-Markovian extensions in fermionic systems). Here, we instead adopt a nonperturbative Green function method to compute the exact NESS for arbitrary values of the system-bath coupling, in order to monitor how the current distribution within the system changes from strong to weak coupling regimes. Interestingly, edge currents appear in the weak-coupling limit but are masked by bulk currents at strong coupling. This contrasts with the common expectation that exotic thermodynamic effects are more prone to arise in strong-coupling configurations~\cite{IlesSmith2014,Esposito2015, Bruch2016,Carrega2016, Strasberg2016, Newman2017, PerarnauLlobet2018,Miller2018,Pancotti2020,Rivas2020,Talkner2020,Popovic2021}. Our exact analysis can also help to identify the range of parameters in which these dissipative edge currents can be experimentally realized. An appropriate way to achieve such realizations is a setup with a high degree of control over microscopic degrees of freedom, e.g.,~quantum simulation on platforms developed to deal with large systems~\cite{Bloch,Szameit,Hafezi2013,EsslingerHaldane,Aidelsburger2015,Schuster, Khanikaev2017, Viyuela2018, TopPhotonReview, Chalopin2020, Esslinger_dissipation, Ferri2021}, whose fundamental constituents may be bosons. 
	
	How can dissipatively robust edge currents arise without band topology? Let us first explore this phenomenon and then we will explain its origin.
	
	\noindent {\em Model.---}We consider the QWZ Hamiltonian~\cite{Qi2006} describing a collection of non-interacting fermions or bosons with two internal ``flavor'' states. The particles occupy a 2D square lattice with sites specified by the coordinates $x = 1,\ldots, L_X$ and $y = 1,\ldots, L_Y$. The Hamiltonian is expressed in terms of vectors of canonical ladder operators ${\bf \a}^\dagger_{x,y} = (\adag_{x,y,\uparrow},\adag_{x,y,\downarrow})$ for each lattice site, whose components create a particle with flavor $\uparrow$ or $\downarrow$. Note that the flavors merely index distinct bands and are unrelated to angular momentum. Explicitly, the Hamiltonian reads as $\H = \H_m + \H_X + \H_Y$, with  ($\hbar=1 = k_B$)
	\begin{align}
		\label{H_m}
		\H_m & = \sum_{x,y}  {\bf \a}^\dagger_{x,y}\cdot (\omega_0\mathbb{1}+ m\sigma_z) \cdot {\bf \a}_{x,y}, \\ 
		\label{H_X}
		\H_X & = \frac{t_X}{2}\sum_{x,y}  {\bf \a}^\dagger_{x+1,y} \cdot  ( \sigma_z + \ii \sigma_y ) \cdot {\bf \a}_{x,y} + {\rm h.c.}, \\
		\label{H_Y}
		\H_Y & = \frac{ t_Y}{2}\sum_{x,y} {\bf \a}^\dagger_{x,y+1} \cdot  ( \sigma_z + \ii \sigma_x ) \cdot {\bf \a}_{x,y} + {\rm h.c.},
	\end{align}
	where $\sigma_{x,y,z}$ are Pauli matrices in flavor space, $\omega_0$ is the on-site energy, $m$ is the flavor energy splitting, and $t_{X,Y} > 0$ are the tunnelling amplitudes in the $x$ and $y$ directions.

	The QWZ Hamiltonian has two notable symmetries. The first is $\hat{\Pi}\hat{R}_\pi$, a combined $\pi$-rotation about the $z$ axis in real space, $\hat{R}_\pi \hat{\mathbf{a}}_{x,y} \hat{R}_\pi^\dagger = \hat{\mathbf{a}}_{L_X+1-x,L_Y+1-y}$, and flavor space, $\hat{\Pi}\hat{\mathbf{a}}_{x,y}\hat{\Pi}^\dagger = \sigma_z \hat{\mathbf{a}}_{x,y}$. The second symmetry is $\hat{\Pi}\hat{\Theta}\hat{\Sigma}_y$, which combines time reversal, $\hat{\Theta}\hat{H}\hat{\Theta}^{-1} = \hat{H}^*$, spatial reflection about the $y$ axis, $\hat{\Sigma}_y\hat{\mathbf{a}}_{x,y}\hat{\Sigma}_y^\dagger = \hat{\mathbf{a}}_{L_X+1-x,y}$, and the flavor $\pi$-rotation $\hat{\Pi}$ defined above. We can already anticipate that these discrete symmetries will play an important role in stabilising edge currents out of equilibrium, as previously found in the context of a bosonic Hofstadter model~\cite{Rivas2017}. Unlike that model, however, the QWZ Hamiltonian exhibits both topologically trivial and nontrivial phases depending on the value of $m$ relative to $t_{X,Y}$. Nontrivial topology manifests as a series of edge states with linear dispersion relation connecting the two single-particle energy bands. Conversely, in the topologically trivial regime there are no edge states and the bands are separated by a finite energy gap. The phase diagram and corresponding band structure is indicated in Fig.~\ref{fig:QWZ}. 
	
	To study an out-of-equilibrium situation, we couple one side of the lattice ($x=1$) to a hot thermal bath at temperature $T_h = \beta^{-1}_h$ and the other side ($x=L_X$) to a cold bath at temperature $T_c = \beta_c^{-1}$, as depicted in Fig.~\ref{fig:QWZ}(b). These baths are modelled by reservoirs of non-interacting fermions or bosons, which can tunnel to and from the system via a linear coupling. We assume that reservoir modes coupled to distinct sites of the system are uncorrelated, and are initially populated according to the distribution function $\bar{n}_{h,c}(\omega) = (\ee^{\beta_{h,c}(\omega-\mu)} \pm 1)^{-1}$, where the plus (minus) sign pertains to fermions (bosons). In the fermionic case, we include a chemical potential $\mu$ to fix the average density, while in the bosonic case we set $\mu=0$. At long times, the system reaches a NESS, which can be computed exactly~\cite{SM}. The NESS is a Gaussian state and thus fully characterized by its correlation matrix $C_{jk} = \langle \adag_k \a_j\rangle$, where the indices $j,k$ represent the coordinates $(x,y)$ as well as the flavor state. Explicitly, we have
	\begin{equation}\label{C_jk}
		\mathbf{C} = \int\frac{\dd\omega}{2\pi} \, \mathbf{G}(\omega) \cdot \Big [\mathbf{ \Gamma}_h \bar{n}_h(\omega) + \mathbf{\Gamma}_c \bar{n}_c(\omega)\Big]\cdot \mathbf{G}^\dagger(\omega),
	\end{equation}
	where $\mathbf{G}(\omega)$ is the retarded Green function of the system obtained by tracing over the reservoirs and $\mathbf{\Gamma}_{c,h}$ are self-energies describing the system-reservoir coupling \cite{Dhar2012,Ryndyk2016}. We work in the wide-band limit, where the self-energies can be approximated by a frequency-independent constant $\gamma$, which we assume to be equal for both hot and cold reservoirs. This approximation, which is valid so long as the reservoir spectral densities vary slowly in the relevant frequency range, significantly simplifies the calculations but is not essential for our conclusions to hold.
	
	The applied thermal gradient gives rise to particle currents flowing within the system. We denote by $J^X_{x,y}$ the mean particle current flowing from site $(x,y)$ to site $(x+1,y)$, while $J^Y_{x,y}$ denotes the current flowing from $(x,y)$ to $(x,y+1)$. These are expectation values of one-body observables and can be found from the NESS correlation matrix $\mathbf{C}$~\cite{SM}.
	
	%\section{Results}
	%\label{sec:results}
	\begin{figure*}
		\begin{minipage}{0.31\linewidth}
			\includegraphics[width=\linewidth]{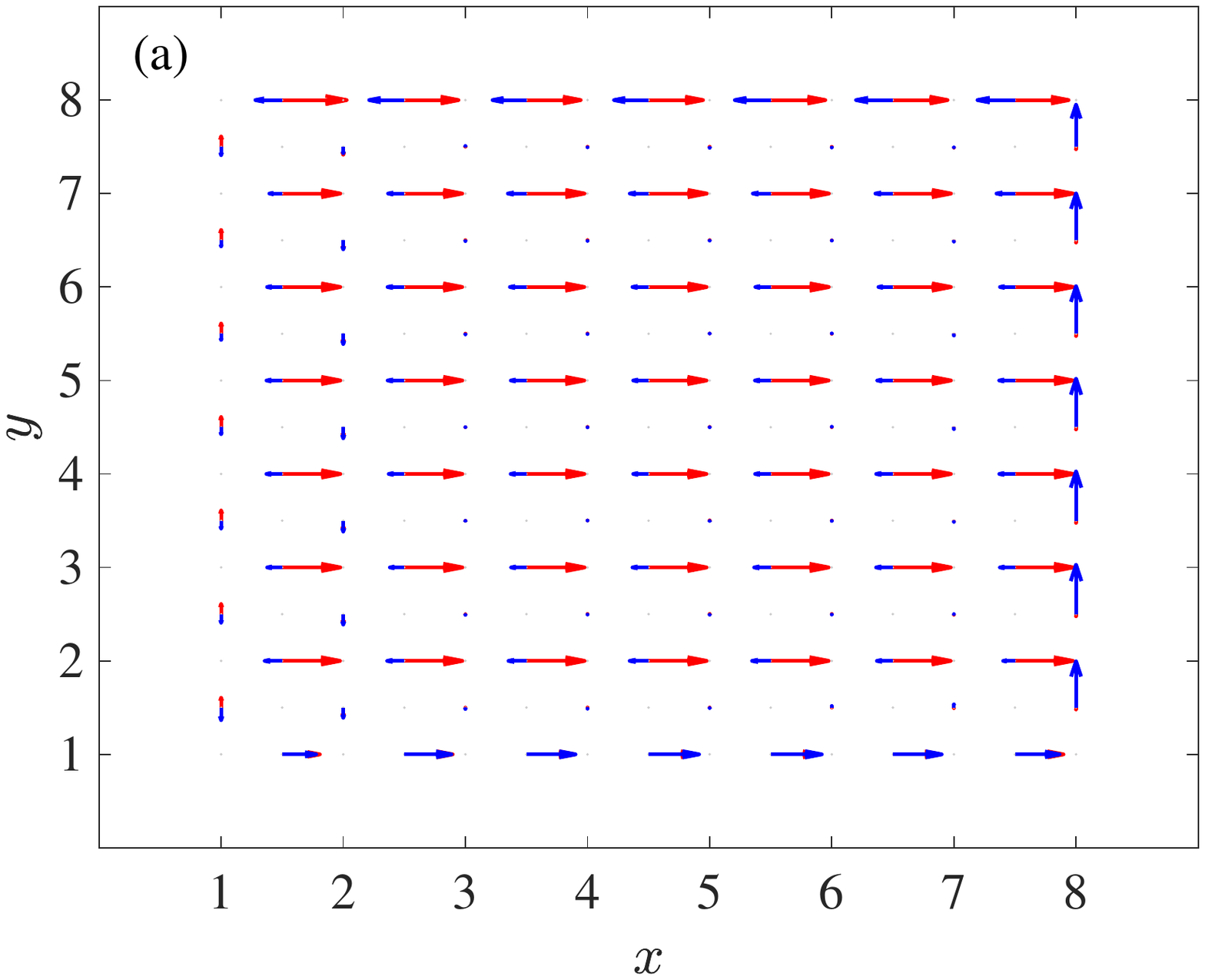}
		\end{minipage}\quad
		\begin{minipage}{0.31\linewidth}
			\includegraphics[width=\linewidth]{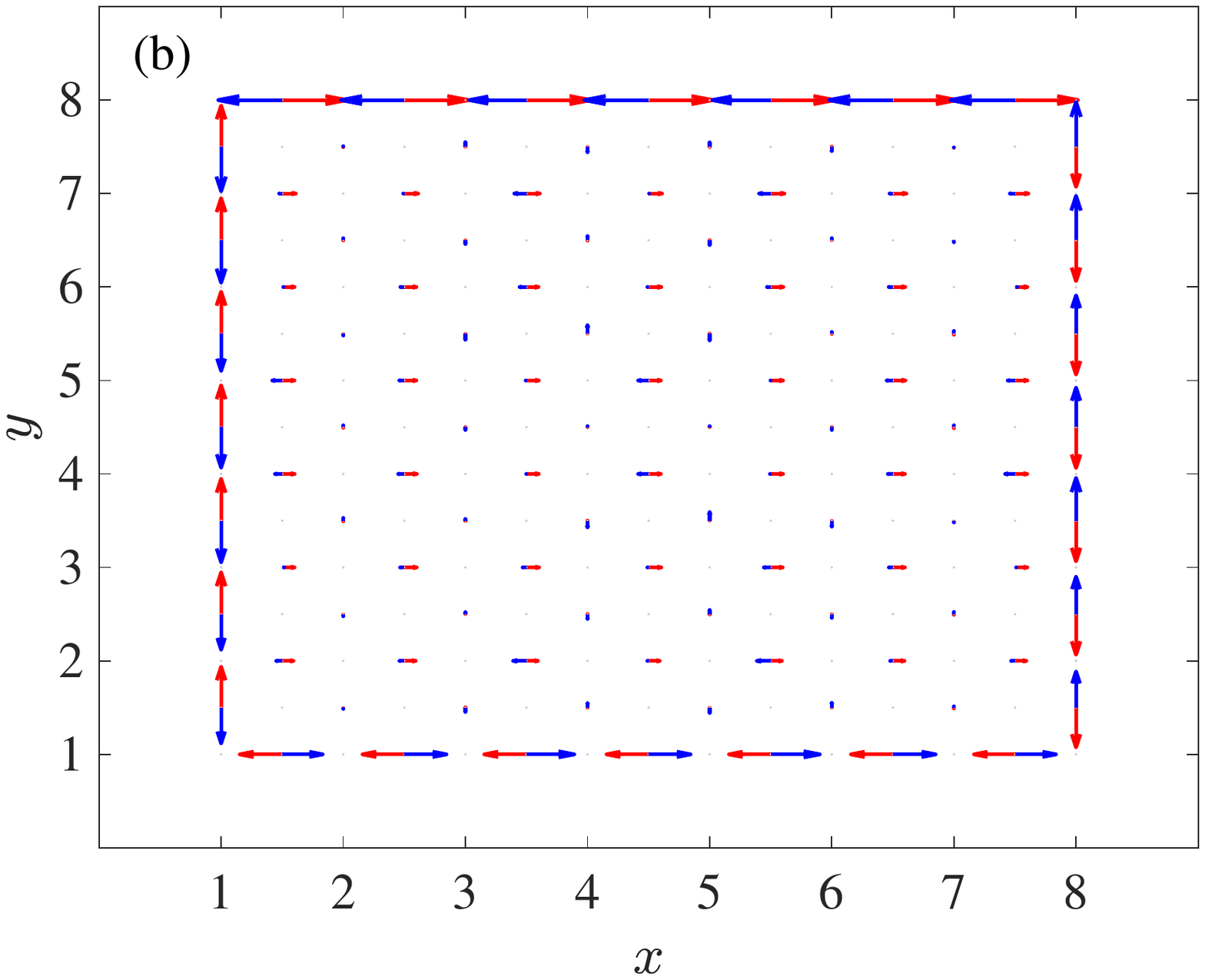}
		\end{minipage}\quad
		\begin{minipage}{0.31\linewidth}
			\includegraphics[width=\linewidth]{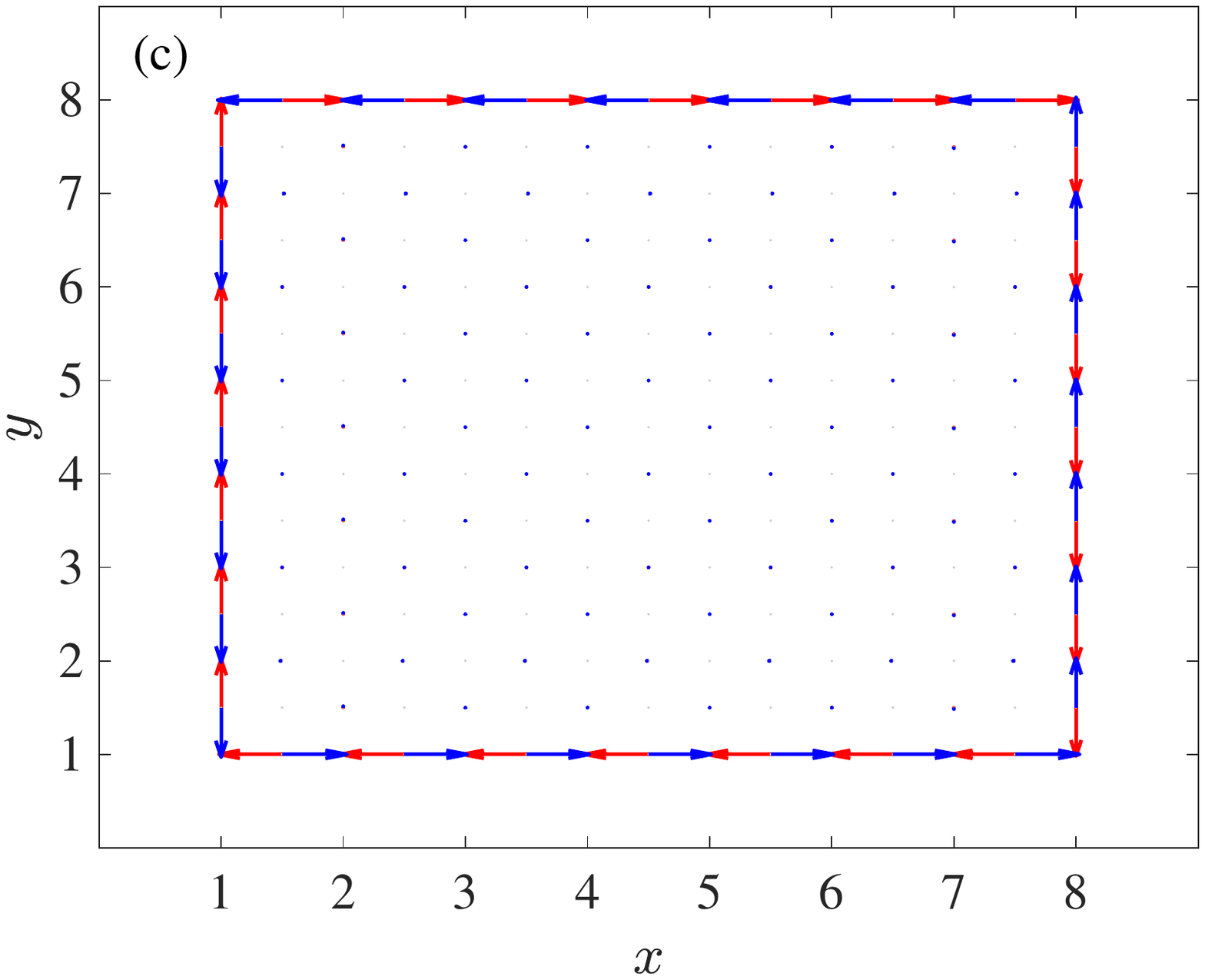}
		\end{minipage} 
		\caption{Current distributions on a lattice of size $L = 8$. Red arrows show boson currents in a topologically trivial phase ($m=3t$), blue arrows show fermion currents in a topologically nontrivial regime ($m=t$). Arrow sizes indicate the magnitude of the currents scaled relative to the largest value within each plot. (a,b) Nonequilibrium current profile with different temperatures $T_h =  t$ and $T_c= 0.01t$, on-site energy $\omega_0=10t$, chemical potential $\mu = \omega_0-0.01t$ for fermions (we always set $\mu=0$ for bosons), and coupling strength (a)~$\gamma = 0.5t$ and (b)~$ \gamma = 0.005t$. (c)~Equilibrium case with temperatures $T_c=T_h=t$, weak coupling, $\gamma = 0.005t$, and other parameters identical to (a,b). \label{fig:quiver_coupling}}
	\end{figure*}
	
	\noindent {\em Results.---}In the following examples, we focus on the symmetric case with $L_X = L_Y = L$ and $t_X = t_Y=t$, so that $|m|<2t$ defines the topologically non-trivial phase. We also fix $\omega_0=10t$ and consider relatively low temperatures, $T_{c,h}\lesssim t$, to accentuate the role of band topology and particle exchange statistics. 
	
	\begin{figure}[b]
		\begin{minipage}{0.495\linewidth}
			\includegraphics[width=\linewidth]{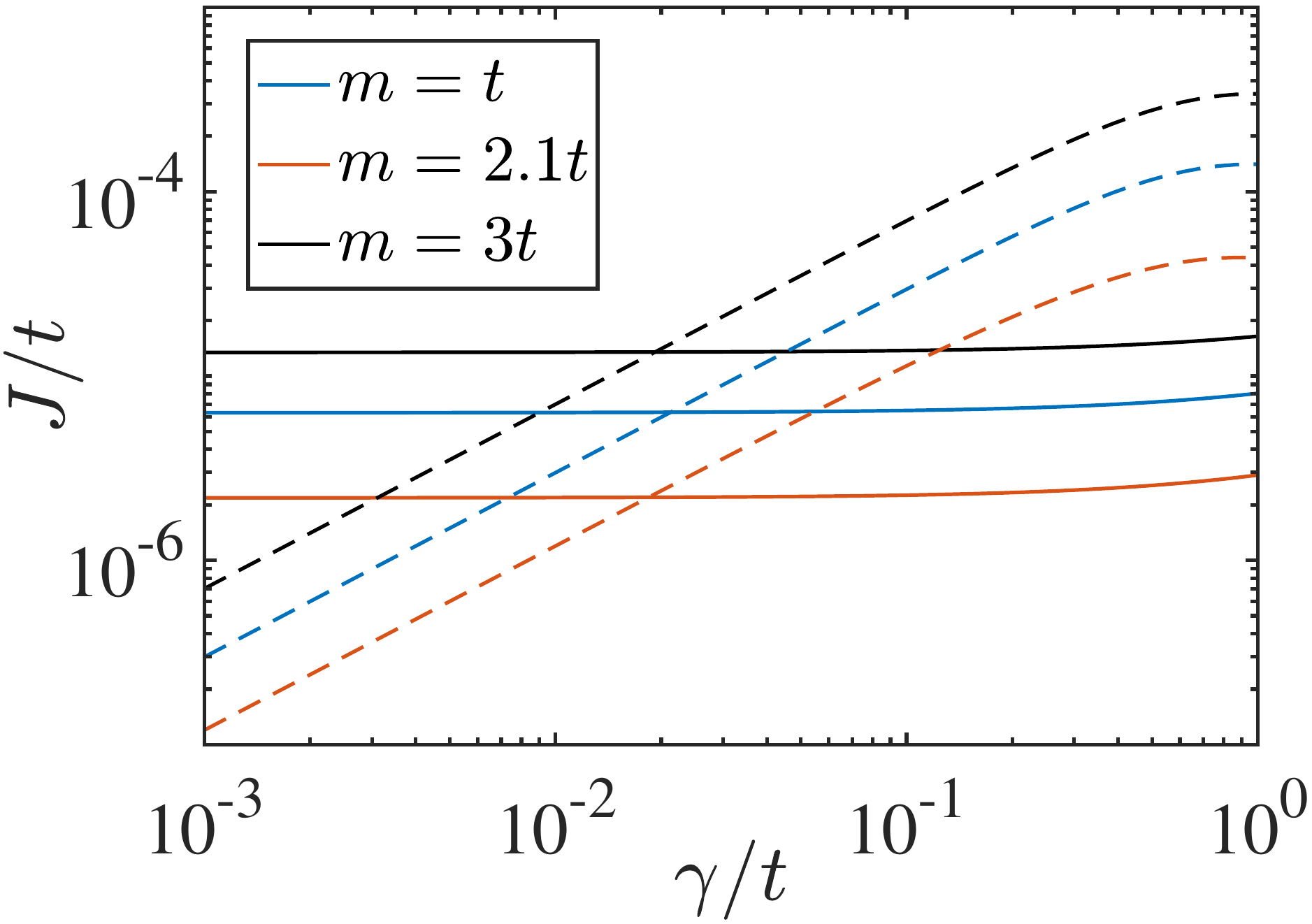}
		\end{minipage}
		\begin{minipage}{0.495\linewidth}
			\includegraphics[width=\linewidth]{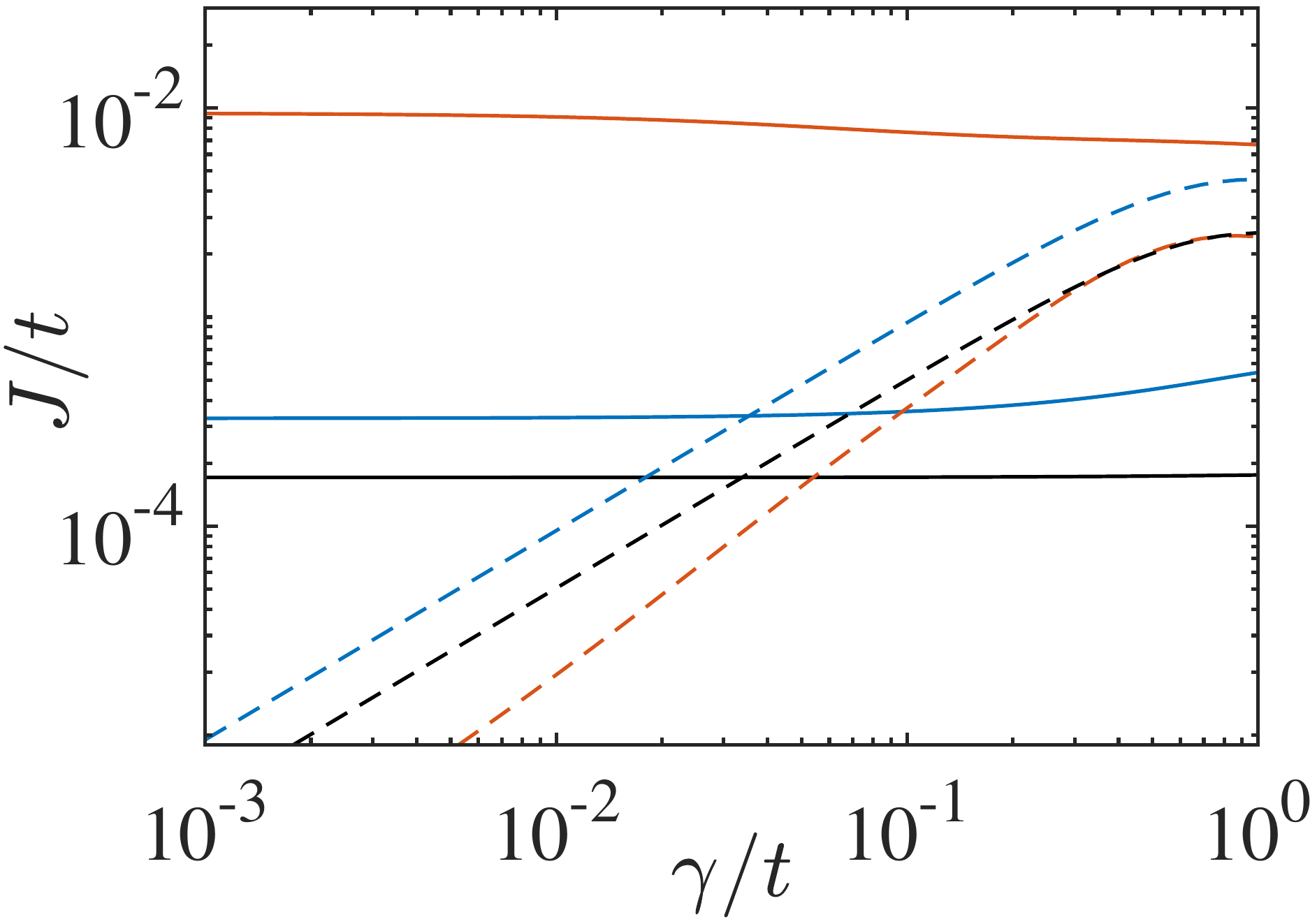}
		\end{minipage}
		\caption{Average edge currents, $J_{\rm edge}$ (solid lines), and bulk currents, $J_{\rm bulk}$ (dashed lines), as a function of the system-bath coupling strength, $\gamma$, for bosons (left panel) and fermions (right panel) in topologically non-trivial ($m/t=1$) and trivial ($m/t=2.1,3$) regimes, and with $T_h= t$, $T_c=0.01t$, $\omega_0=10t$, and $\mu = \omega_0+0.1t$. \label{fig:edge_bulk_currents}}
	\end{figure}
	
	Figs.~\ref{fig:quiver_coupling}(a,b) plot the nonequilibrium current distributions for two different values of the system-reservoir coupling, $\gamma$. Red arrows show the currents for a bosonic system with $m=3t$ (similar results are obtained for $|m|<2t$). Remarkably, the currents become progressively localized on the boundary of the system as $\gamma$ is reduced, even though the band topology is trivial. These edge currents also arise in the equilibrium case, $T_h=T_c$, as shown in Fig.~\ref{fig:quiver_coupling}(c). Qualitatively similar results are obtained for fermions, as shown for a nontrivial phase ($m=t$) by the blue arrows in Fig.~\ref{fig:quiver_coupling}. The direction of fermionic particle flow depends on the chemical potential; for the parameters in Figs.~\ref{fig:quiver_coupling}(a,b), the thermoelectric induced current flows in the opposite direction to the temperature gradient. Conversely, when $\mu=\omega_0$, all currents vanish due to particle-hole symmetry~\cite{SM}.
	
	In order to quantify the emergence of boundary currents more precisely, we define the average edge and bulk currents
	\begin{align}\label{edge_and_bulk_current}
		J_{\rm edge} &= \frac{1}{2L_X}\sum_{x} \left( J^{X}_{x,L_Y} - J^X_{x,1} \right ), 	\qquad J_{\rm bulk} = \frac{1}{L_Y}\sum_{y} J^{X}_{x,y}.
	\end{align}
	The choice of $x$-coordinate in the definition of $J_{\rm bulk}$ is arbitrary due to particle-number conservation; we take $x=\lfloor L_X/2\rfloor$. The total current flowing between the two reservoirs is given by $J_{\rm tot} = L_Y J_{\rm bulk}$. 
	
	We plot the edge and bulk currents in Fig.~\ref{fig:edge_bulk_currents} for bosonic and fermionic systems as a function of the system-reservoir coupling strength, $\gamma$. In the bosonic case, both bulk and edge currents increase with $m$ and are thus larger in the topologically trivial phase. The situation is reversed for fermions, with the largest edge currents obtained in the topologically nontrivial phase. In all cases, the bulk currents are proportional to $\gamma$ and thus vanish as $\gamma\to 0$, while the edge currents remains invariant for a wide range of values of $\gamma$ and persist in the weak-coupling limit. These quantitative results therefore confirm the qualitative picture of Fig.~\ref{fig:quiver_coupling}, i.e., edge currents arise for small coupling irrespectively of band topology or particle statistics. In topologically trivial phases, edge currents begin to dominate once $\gamma$ becomes comparable to the level spacing of the single-particle Hamiltonian. However, in the topologically nontrivial phase for fermions near half filling,  currents remain localized on the edges even for strong coupling, $\gamma\sim  t$. 
	
	\begin{figure*}
		\begin{minipage}{0.31\linewidth}
			\includegraphics[width=\linewidth]{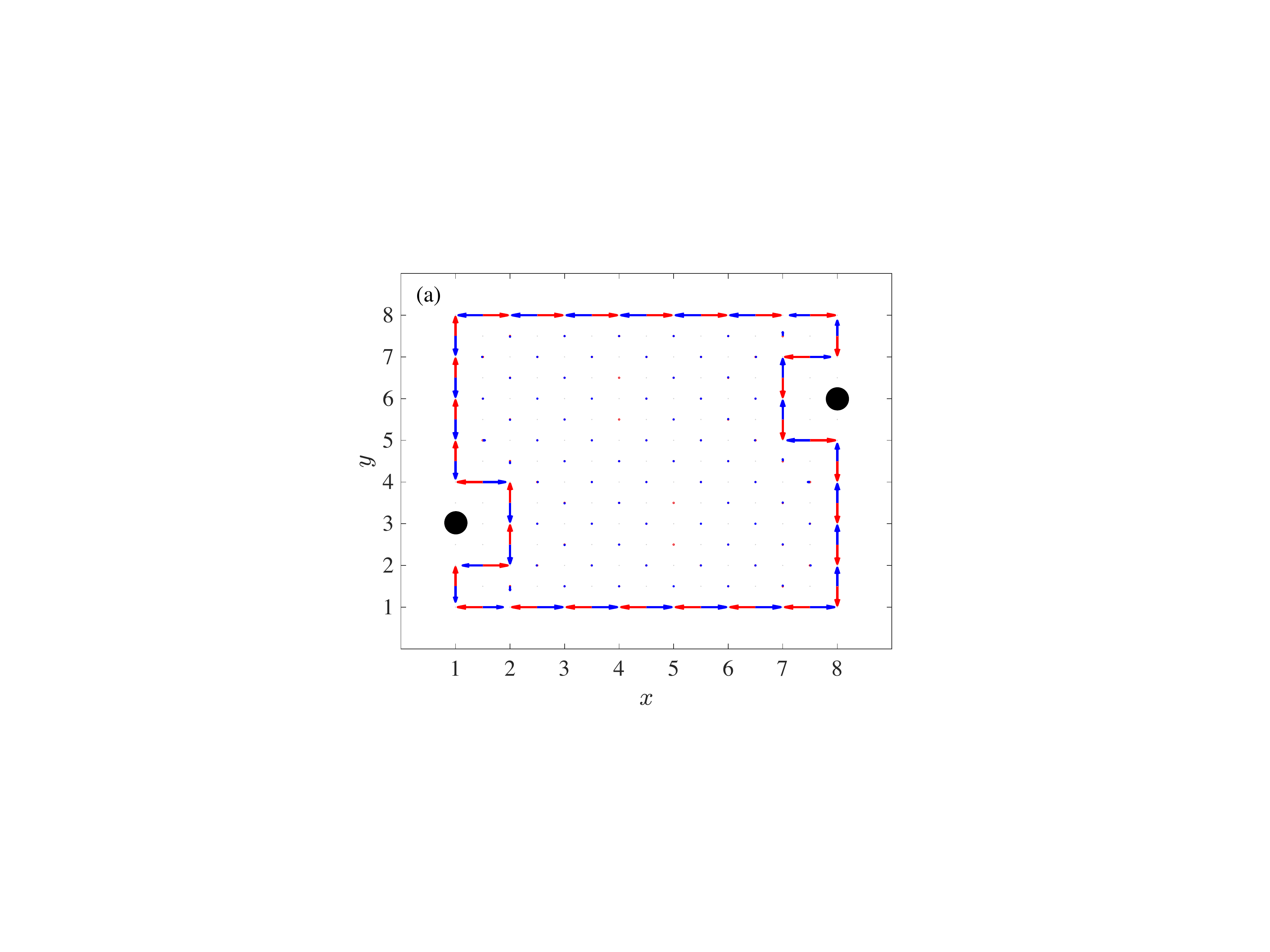}
		\end{minipage}\quad
		\begin{minipage}{0.31\linewidth}
			\includegraphics[width=\linewidth]{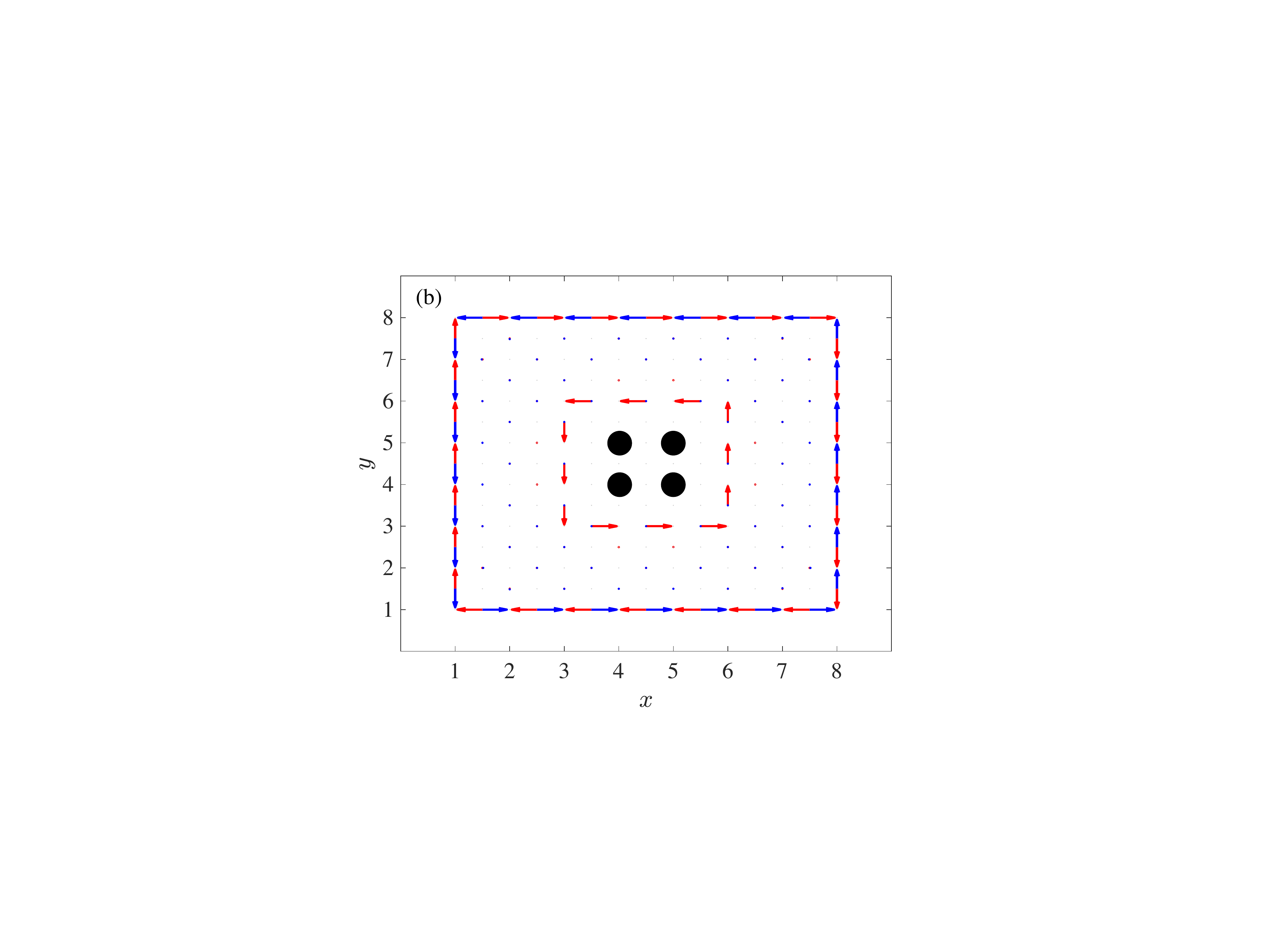}
		\end{minipage} \quad
		\begin{minipage}{0.31\linewidth}
			\includegraphics[width=\linewidth]{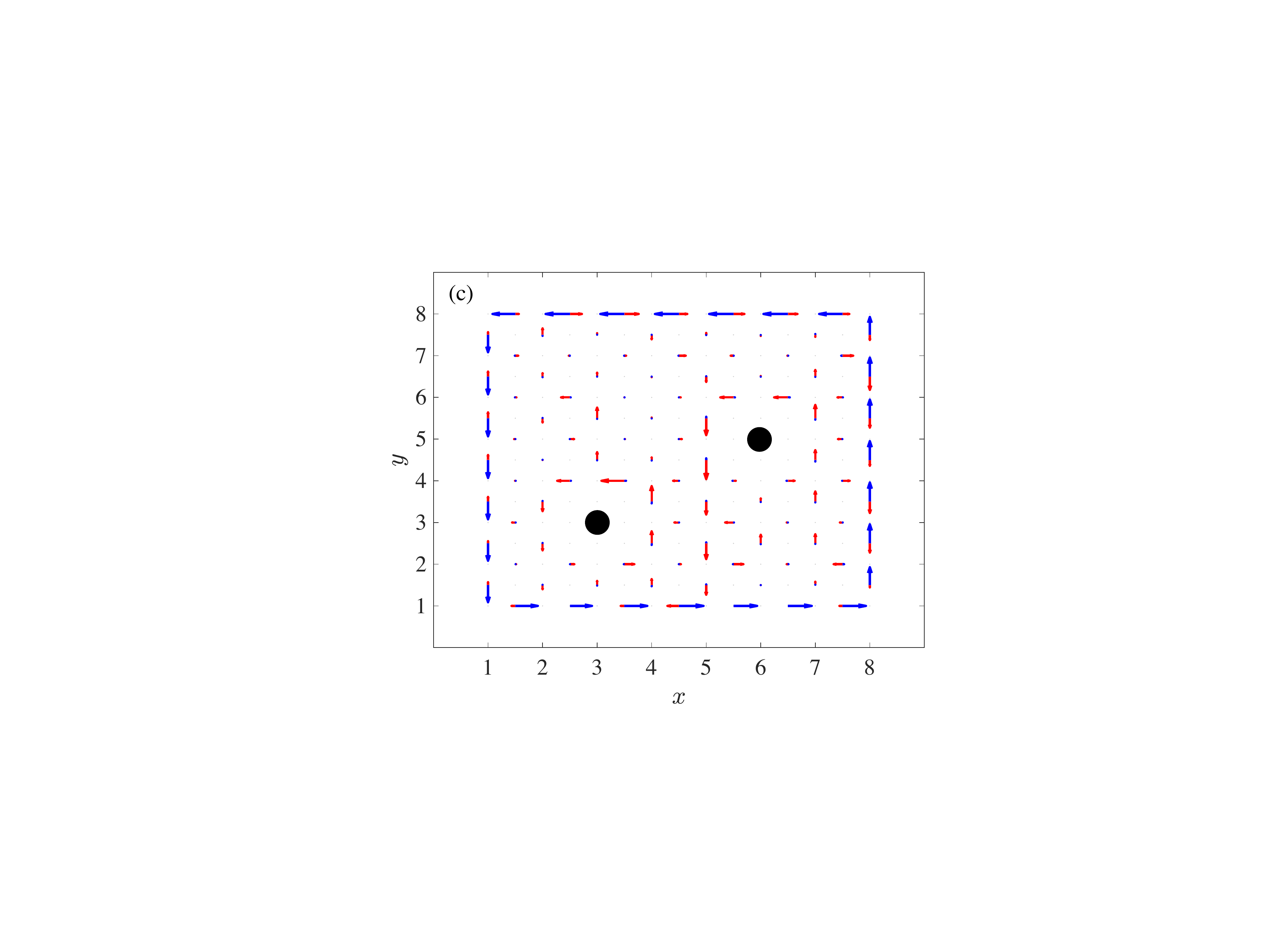}
		\end{minipage}
		\caption{Effect of impurities on the current distribution in the weak-coupling limit, with impurity positions marked by filled black circles. Red arrows show boson currents and blue arrows show fermion currents. In all plots, $m=t$, $T_h= t$, $T_c=0.01t$, $\omega_0=10t$ and $\mu = \omega_0 - 0.1t$.\label{fig:quiver_impurities}}
	\end{figure*}

	Finally, we investigate whether these effects withstand the presence of imperfections, a property which defines the notion of dissipatively robust currents. Specifically, we examine how the currents change when static impurities are added to the lattice. Impurities are modelled by a large on-site energy shift, i.e.,~a term $\Delta {\bf\a}^\dagger_{x,y}
	\cdot {\bf \a}_{x,y}$ added to the Hamitonian, where $\Delta \gg \omega_0, t$ and $(x,y)$ are the coordinates of the impurity site~\cite{Rivas2017}. We focus hereafter on the weak-coupling limit where the edge currents are most prominent in the absence of impurities. In this limit, an analytical expression for the NESS~\cite{Dhar2012} can be obtained, which is identical to the solution of the Lindblad equation derived under the Born-Markov and secular approximations~\cite{SM}.
	
	In Fig.~\ref{fig:quiver_impurities} we show three examples of how impurities affect the nonequilibrium current distribution in the topologically nontrivial phase. We observe substantial differences between bosonic and fermionic systems when one of the reservoirs is at low temperature. The bosonic edge currents are preserved only when the impurity distribution is invariant under one of the symmetries~$\hat{\Pi}\hat{\Theta}\hat{\Sigma}_y$ or $\hat{\Pi}\hat{R}_\pi$, which correspond to the purely spatial symmetries $\hat{\Sigma}_y$ and $\hat{R}_\pi$ since the defects satisfy $\hat{\Pi}$ and $\hat{\Theta}$ automatically. In the case of impurities placed on the edge (and assuming the symmetries are respected in the bosonic case), the currents simply detour around the impurity sites [Fig.~\ref{fig:quiver_impurities}(a)]. If impurities are placed in the bulk of a bosonic system in an appropriately symmetric way, counter-currents shield the impurity sites by circulating in the opposite direction to the edge currents~[Fig.~\ref{fig:quiver_impurities}(b)]. In the absence of these symmetries, the bosonic edge currents are strongly disrupted [Fig.~\ref{fig:quiver_impurities}(c)]. For fermions in a topological phase, the situation is starkly different when at least one reservoir is at low temperature: impurities placed in the bulk have no effect whatsoever on the current distribution~[Figs.~\ref{fig:quiver_impurities}(b,c)]. In the topologically trivial phase or at high temperature, both bosonic and fermionic edge currents enjoy the same symmetry-protected robustness as  bosons in the nontrivial phase. 
	
\noindent {\em Discussion.---}The question posed in the introduction on the nontopological origin of the edge currents can now be answered. In the weak-coupling limit, nonequilibrium coherences between energy eigenstates become small~\cite{Dhar2012} and $J_{\rm bulk}$, which is directly proportional to these coherences~\cite{Mitchison2018}, is negligible in comparison to the contributions from individual energy eigenstates. Moreover, the single-particle eigenmodes in the QWZ model present a nonzero Berry curvature $\bm{\mathcal{F}}_\alpha$ for $m\neq 0$~\cite{Qi2006}, which breaks time-reversal symmetry and endows the eigenmodes with a particular chirality. This can be understood in a semi-classical picture~\cite{Haldane2004}, where a wavepacket propagates with velocity $\mathbf{v}_\alpha = \partial_{\mathbf{k}} \omega_\alpha +  \nabla V \times \bm{\mathcal{F}}_\alpha$, with $\mathbf{k}$ the wavevector and $V(x,y)$ a confining potential describing the edge of the system~\cite{Xiao2010}. Assuming that the non-equilibrium distribution function only depends on energy, the net current due to the derivative term $\partial_{\mathbf{k}} \omega_\alpha$ can be shown to vanish in the bulk due to cancellations between positive and negative wavevectors~\cite{SM}. However, the contribution of the axial force $\nabla V \times \bm{\mathcal{F}}_\alpha$ is non-zero near the boundary, thus generating a net chiral edge current~\cite{SM}. This erasure effect arises whenever there is broken time-reversal symmetry and a smooth, energetically monotonic distribution of chiral eigenmodes~\cite{Rivas2017}; it is thus independent of exchange statistics or band topology. Intuitively, the effect can be understood in terms of circulating currents that cancel in the bulk but add constructively on the boundary [Fig.~\ref{fig:QWZ}(b)].
	
A chiral current between two thermal reservoirs unavoidably entails a local cross-current phenomenon:  on one edge there are particles flowing from the cold bath to the hot bath. In bosonic systems, this implies a ``violation'' of the second law of thermodynamics within a one-dimensional subsystem, i.e.,~energy flows against the temperature gradient along one edge even though the total rate of entropy production is positive. This effect was reported in Refs.~\cite{Rivas2017,Xing2020} for bosonic lattices governed by the Hofstadter Hamiltonian. Now, we see that the key requisite for this behavior is not the nontrivial band topology of the lattice, but rather the nonzero Berry curvature in a regime of weak system-bath coupling [Fig.~\ref{fig:quiver_coupling}(b,c)].
	
The edge currents are robust to impurities if the defect distribution satisfies either of the symmetries $\hat{\Pi}\hat{\Theta}\hat{\Sigma}_y$ or $\hat{\Pi}\hat{R}_\pi$. Note that the relevant nonequilibrium symmetries are determined both by the Hamiltonian and by the configuration of the baths. These symmetries leave the NESS invariant and lead to a steady-state correlation matrix that is independent of the spatial orientation of the reservoirs. Under these conditions, the nonequilibrium distribution function is simply the average of the reservoir distributions~\cite{SM}, i.e.,~$n(\omega_\alpha) = \tfrac{1}{2}\left [ \bar{n}_h(\omega_\alpha) + \bar{n}_c(\omega_\alpha)\right ]$ is the population of the eigenmode with frequency $\omega_\alpha$. This smooth distribution function yields boundary currents due to the erasure effect in the bulk. In the absence of symmetry, however, the contribution of each mode depends not only on energy but also on the spatial profile of the corresponding wavefunction. This creates an erratically varying distribution function that destroys the erasure effect~\cite{SM}. 

For fermions at low temperature $T_c$, $n(\omega_\alpha$) undergoes a sharp change near $\omega_\alpha \approx \mu$. In a topologically nontrivial phase near half filling, this feature acts as a filter that populates only one or two edge states~\cite{SM}, which are topologically protected against perturbations in the bulk. The corresponding boundary current is thus completely unaffected by impurities [Figs.~\ref{fig:quiver_impurities}(b,c)]. Remarkably, this holds for arbitrarily large $T_h$, providing an instance of a topologically protected property that is also stable under dissipation. The number and character of edge modes that contribute to the current pattern is determined by the chemical potential~\cite{SM}.
	
In summary, our nonperturbative analysis has shown that robust boundary currents emerge at weak coupling in the QWZ lattice driven out of equilibrium by a thermal gradient. Our results reveal a novel dissipative mechanism for symmetry-protected edge transport, which arises in bosonic and fermionic systems with and without nontrivial band topology. This makes quantum simulators based on either fermionic or bosonic degrees of freedom attractive candidates to observe this exotic boundary physics.
	
\textit{Acknowledgements.---}M.~T.~M. acknowledges funding from the ERC Starting  Grant ODYSSEY (Grant  Agreement No.~758403) and the EPSRC-SFI Joint Funding of Research project QuamNESS. A.~R. and M.~A.~M.-D. acknowledge financial support from the Spanish MINECO grants MINECO/FEDER Projects FIS2017-91460-EXP, PGC2018-099169-B-I00 FIS-2018 and from CAM/FEDER Project No. S2018/TCS-4342 (QUITEMAD-CM). The research of A.~R. and M.~A.~M.-D. has been partially supported by the U.S. Army Research Office through Grant No.~W911NF-14-1-0103. Calculations were performed on the Lonsdale cluster maintained by the Trinity Centre for High Performance Computing. This cluster was funded through grants from SFI. 
	
	\bibliographystyle{apsrev4-2}
	\bibliography{bibliography}

\begin{thebibliography}{62}%
\makeatletter
\providecommand \@ifxundefined [1]{%
 \@ifx{#1\undefined}
}%
\providecommand \@ifnum [1]{%
 \ifnum #1\expandafter \@firstoftwo
 \else \expandafter \@secondoftwo
 \fi
}%
\providecommand \@ifx [1]{%
 \ifx #1\expandafter \@firstoftwo
 \else \expandafter \@secondoftwo
 \fi
}%
\providecommand \natexlab [1]{#1}%
\providecommand \enquote  [1]{``#1''}%
\providecommand \bibnamefont  [1]{#1}%
\providecommand \bibfnamefont [1]{#1}%
\providecommand \citenamefont [1]{#1}%
\providecommand \href@noop [0]{\@secondoftwo}%
\providecommand \href [0]{\begingroup \@sanitize@url \@href}%
\providecommand \@href[1]{\@@startlink{#1}\@@href}%
\providecommand \@@href[1]{\endgroup#1\@@endlink}%
\providecommand \@sanitize@url [0]{\catcode `\\12\catcode `\$12\catcode
  `\&12\catcode `\#12\catcode `\^12\catcode `\_12\catcode `\%12\relax}%
\providecommand \@@startlink[1]{}%
\providecommand \@@endlink[0]{}%
\providecommand \url  [0]{\begingroup\@sanitize@url \@url }%
\providecommand \@url [1]{\endgroup\@href {#1}{\urlprefix }}%
\providecommand \urlprefix  [0]{URL }%
\providecommand \Eprint [0]{\href }%
\providecommand \doibase [0]{https://doi.org/}%
\providecommand \selectlanguage [0]{\@gobble}%
\providecommand \bibinfo  [0]{\@secondoftwo}%
\providecommand \bibfield  [0]{\@secondoftwo}%
\providecommand \translation [1]{[#1]}%
\providecommand \BibitemOpen [0]{}%
\providecommand \bibitemStop [0]{}%
\providecommand \bibitemNoStop [0]{.\EOS\space}%
\providecommand \EOS [0]{\spacefactor3000\relax}%
\providecommand \BibitemShut  [1]{\csname bibitem#1\endcsname}%
\let\auto@bib@innerbib\@empty
%</preamble>
\bibitem [{\citenamefont {Haldane}(1988)}]{Haldane}%
  \BibitemOpen
  \bibfield  {author} {\bibinfo {author} {\bibfnamefont {F.~D.~M.}\
  \bibnamefont {Haldane}},\ }\href
  {https://doi.org/10.1103/PhysRevLett.61.2015} {\bibfield  {journal} {\bibinfo
   {journal} {Phys. Rev. Lett.}\ }\textbf {\bibinfo {volume} {61}},\ \bibinfo
  {pages} {2015} (\bibinfo {year} {1988})}\BibitemShut {NoStop}%
\bibitem [{\citenamefont {Kane}\ and\ \citenamefont {Mele}(2005)}]{KaneMele}%
  \BibitemOpen
  \bibfield  {author} {\bibinfo {author} {\bibfnamefont {C.~L.}\ \bibnamefont
  {Kane}}\ and\ \bibinfo {author} {\bibfnamefont {E.~J.}\ \bibnamefont
  {Mele}},\ }\href {https://doi.org/10.1103/PhysRevLett.95.226801} {\bibfield
  {journal} {\bibinfo  {journal} {Phys. Rev. Lett.}\ }\textbf {\bibinfo
  {volume} {95}},\ \bibinfo {pages} {226801} (\bibinfo {year}
  {2005})}\BibitemShut {NoStop}%
\bibitem [{\citenamefont {Bernevig}\ \emph {et~al.}(2006)\citenamefont
  {Bernevig}, \citenamefont {Hughes},\ and\ \citenamefont {Zhang}}]{Bernevig}%
  \BibitemOpen
  \bibfield  {author} {\bibinfo {author} {\bibfnamefont {B.~A.}\ \bibnamefont
  {Bernevig}}, \bibinfo {author} {\bibfnamefont {T.~L.}\ \bibnamefont
  {Hughes}},\ and\ \bibinfo {author} {\bibfnamefont {S.-C.}\ \bibnamefont
  {Zhang}},\ }\href {https://doi.org/10.1126/science.1133734} {\bibfield
  {journal} {\bibinfo  {journal} {Science}\ }\textbf {\bibinfo {volume}
  {314}},\ \bibinfo {pages} {1757} (\bibinfo {year} {2006})}\BibitemShut
  {NoStop}%
\bibitem [{\citenamefont {K{\"o}nig}\ \emph {et~al.}(2007)\citenamefont
  {K{\"o}nig}, \citenamefont {Wiedmann}, \citenamefont {Br{\"u}ne},
  \citenamefont {Roth}, \citenamefont {Buhmann}, \citenamefont {Molenkamp},
  \citenamefont {Qi},\ and\ \citenamefont {Zhang}}]{ExpZhang}%
  \BibitemOpen
  \bibfield  {author} {\bibinfo {author} {\bibfnamefont {M.}~\bibnamefont
  {K{\"o}nig}}, \bibinfo {author} {\bibfnamefont {S.}~\bibnamefont {Wiedmann}},
  \bibinfo {author} {\bibfnamefont {C.}~\bibnamefont {Br{\"u}ne}}, \bibinfo
  {author} {\bibfnamefont {A.}~\bibnamefont {Roth}}, \bibinfo {author}
  {\bibfnamefont {H.}~\bibnamefont {Buhmann}}, \bibinfo {author} {\bibfnamefont
  {L.~W.}\ \bibnamefont {Molenkamp}}, \bibinfo {author} {\bibfnamefont {X.-L.}\
  \bibnamefont {Qi}},\ and\ \bibinfo {author} {\bibfnamefont {S.-C.}\
  \bibnamefont {Zhang}},\ }\href {https://doi.org/10.1126/science.1148047}
  {\bibfield  {journal} {\bibinfo  {journal} {Science}\ }\textbf {\bibinfo
  {volume} {318}},\ \bibinfo {pages} {766} (\bibinfo {year}
  {2007})}\BibitemShut {NoStop}%
\bibitem [{\citenamefont {Qi}\ and\ \citenamefont {Zhang}(2011)}]{QiZhang}%
  \BibitemOpen
  \bibfield  {author} {\bibinfo {author} {\bibfnamefont {X.-L.}\ \bibnamefont
  {Qi}}\ and\ \bibinfo {author} {\bibfnamefont {S.-C.}\ \bibnamefont {Zhang}},\
  }\href {https://doi.org/10.1103/RevModPhys.83.1057} {\bibfield  {journal}
  {\bibinfo  {journal} {Rev. Mod. Phys.}\ }\textbf {\bibinfo {volume} {83}},\
  \bibinfo {pages} {1057} (\bibinfo {year} {2011})}\BibitemShut {NoStop}%
\bibitem [{\citenamefont {Hasan}\ and\ \citenamefont {Kane}(2010)}]{Hasan}%
  \BibitemOpen
  \bibfield  {author} {\bibinfo {author} {\bibfnamefont {M.~Z.}\ \bibnamefont
  {Hasan}}\ and\ \bibinfo {author} {\bibfnamefont {C.~L.}\ \bibnamefont
  {Kane}},\ }\href {https://doi.org/10.1103/RevModPhys.82.3045} {\bibfield
  {journal} {\bibinfo  {journal} {Rev. Mod. Phys.}\ }\textbf {\bibinfo {volume}
  {82}},\ \bibinfo {pages} {3045} (\bibinfo {year} {2010})}\BibitemShut
  {NoStop}%
\bibitem [{\citenamefont {Breuer}\ and\ \citenamefont
  {Petruccione}(2002)}]{Breuer2002}%
  \BibitemOpen
  \bibfield  {author} {\bibinfo {author} {\bibfnamefont {H.-P.}\ \bibnamefont
  {Breuer}}\ and\ \bibinfo {author} {\bibfnamefont {F.}~\bibnamefont
  {Petruccione}},\ }\href
  {https://doi.org/10.1093/acprof:oso/9780199213900.001.0001} {\emph {\bibinfo
  {title} {The theory of open quantum systems}}}\ (\bibinfo  {publisher}
  {Oxford University Press},\ \bibinfo {address} {Oxford},\ \bibinfo {year}
  {2002})\BibitemShut {NoStop}%
\bibitem [{\citenamefont {Gardiner}\ and\ \citenamefont
  {Zoller}(2004)}]{GardinerZoller}%
  \BibitemOpen
  \bibfield  {author} {\bibinfo {author} {\bibfnamefont {C.}~\bibnamefont
  {Gardiner}}\ and\ \bibinfo {author} {\bibfnamefont {P.}~\bibnamefont
  {Zoller}},\ }\href {https://www.springer.com/book/9783540223016} {\emph
  {\bibinfo {title} {Quantum noise: a handbook of Markovian and non-Markovian
  quantum stochastic methods with applications to quantum optics}}}\ (\bibinfo
  {publisher} {Springer},\ \bibinfo {address} {Berlin},\ \bibinfo {year}
  {2004})\BibitemShut {NoStop}%
\bibitem [{\citenamefont {Kamenev}(2009)}]{Kamenev2009}%
  \BibitemOpen
  \bibfield  {author} {\bibinfo {author} {\bibfnamefont {A.}~\bibnamefont
  {Kamenev}},\ }\href {https://doi.org/10.1017/cbo9781139003667} {\emph
  {\bibinfo {title} {Field Theory of Non-Equilibrium Systems}}}\ (\bibinfo
  {publisher} {Cambridge University Press},\ \bibinfo {address} {Cambridge},\
  \bibinfo {year} {2009})\BibitemShut {NoStop}%
\bibitem [{\citenamefont {Rivas}\ and\ \citenamefont
  {Huelga}(2012)}]{RivasHuelga}%
  \BibitemOpen
  \bibfield  {author} {\bibinfo {author} {\bibfnamefont {A.}~\bibnamefont
  {Rivas}}\ and\ \bibinfo {author} {\bibfnamefont {S.~F.}\ \bibnamefont
  {Huelga}},\ }\href {https://doi.org/10.1007/978-3-642-23354-8} {\emph
  {\bibinfo {title} {Open quantum systems. An introduction}}}\ (\bibinfo
  {publisher} {Springer},\ \bibinfo {address} {Heidelberg},\ \bibinfo {year}
  {2012})\BibitemShut {NoStop}%
\bibitem [{\citenamefont {Landi}\ \emph {et~al.}(2021)\citenamefont {Landi},
  \citenamefont {Poletti},\ and\ \citenamefont {Schaller}}]{Landi2021}%
  \BibitemOpen
  \bibfield  {author} {\bibinfo {author} {\bibfnamefont {G.~T.}\ \bibnamefont
  {Landi}}, \bibinfo {author} {\bibfnamefont {D.}~\bibnamefont {Poletti}},\
  and\ \bibinfo {author} {\bibfnamefont {G.}~\bibnamefont {Schaller}},\
  }\href@noop {} {\bibinfo {title} {Non-equilibrium boundary driven quantum
  systems: models, methods and properties}} (\bibinfo {year} {2021}),\ \Eprint
  {https://arxiv.org/abs/2104.14350} {arXiv:2104.14350 [quant-ph]} \BibitemShut
  {NoStop}%
\bibitem [{\citenamefont {Qi}\ \emph {et~al.}(2006)\citenamefont {Qi},
  \citenamefont {Wu},\ and\ \citenamefont {Zhang}}]{Qi2006}%
  \BibitemOpen
  \bibfield  {author} {\bibinfo {author} {\bibfnamefont {X.-L.}\ \bibnamefont
  {Qi}}, \bibinfo {author} {\bibfnamefont {Y.-S.}\ \bibnamefont {Wu}},\ and\
  \bibinfo {author} {\bibfnamefont {S.-C.}\ \bibnamefont {Zhang}},\ }\href
  {https://doi.org/10.1103/PhysRevB.74.085308} {\bibfield  {journal} {\bibinfo
  {journal} {Phys. Rev. B}\ }\textbf {\bibinfo {volume} {74}},\ \bibinfo
  {pages} {085308} (\bibinfo {year} {2006})}\BibitemShut {NoStop}%
\bibitem [{\citenamefont {Diehl}\ \emph {et~al.}(2011)\citenamefont {Diehl},
  \citenamefont {Rico}, \citenamefont {Baranov},\ and\ \citenamefont
  {Zoller}}]{Diehl2011}%
  \BibitemOpen
  \bibfield  {author} {\bibinfo {author} {\bibfnamefont {S.}~\bibnamefont
  {Diehl}}, \bibinfo {author} {\bibfnamefont {E.}~\bibnamefont {Rico}},
  \bibinfo {author} {\bibfnamefont {M.~A.}\ \bibnamefont {Baranov}},\ and\
  \bibinfo {author} {\bibfnamefont {P.}~\bibnamefont {Zoller}},\ }\href
  {https://doi.org/10.1038/nphys2106} {\bibfield  {journal} {\bibinfo
  {journal} {Nature Phys.}\ }\textbf {\bibinfo {volume} {7}},\ \bibinfo {pages}
  {971} (\bibinfo {year} {2011})}\BibitemShut {NoStop}%
\bibitem [{\citenamefont {Viyuela}\ \emph {et~al.}(2012)\citenamefont
  {Viyuela}, \citenamefont {Rivas},\ and\ \citenamefont
  {Martin-Delgado}}]{Viyuela2012}%
  \BibitemOpen
  \bibfield  {author} {\bibinfo {author} {\bibfnamefont {O.}~\bibnamefont
  {Viyuela}}, \bibinfo {author} {\bibfnamefont {A.}~\bibnamefont {Rivas}},\
  and\ \bibinfo {author} {\bibfnamefont {M.~A.}\ \bibnamefont
  {Martin-Delgado}},\ }\href {https://doi.org/10.1103/PhysRevB.86.155140}
  {\bibfield  {journal} {\bibinfo  {journal} {Phys. Rev. B}\ }\textbf {\bibinfo
  {volume} {86}},\ \bibinfo {pages} {155140} (\bibinfo {year}
  {2012})}\BibitemShut {NoStop}%
\bibitem [{\citenamefont {Rivas}\ \emph {et~al.}(2013)\citenamefont {Rivas},
  \citenamefont {Viyuela},\ and\ \citenamefont {Martin-Delgado}}]{Rivas2013}%
  \BibitemOpen
  \bibfield  {author} {\bibinfo {author} {\bibfnamefont {A.}~\bibnamefont
  {Rivas}}, \bibinfo {author} {\bibfnamefont {O.}~\bibnamefont {Viyuela}},\
  and\ \bibinfo {author} {\bibfnamefont {M.~A.}\ \bibnamefont
  {Martin-Delgado}},\ }\href {https://doi.org/10.1103/PhysRevB.88.155141}
  {\bibfield  {journal} {\bibinfo  {journal} {Phys. Rev. B}\ }\textbf {\bibinfo
  {volume} {88}},\ \bibinfo {pages} {155141} (\bibinfo {year}
  {2013})}\BibitemShut {NoStop}%
\bibitem [{\citenamefont {Budich}\ \emph {et~al.}(2015)\citenamefont {Budich},
  \citenamefont {Zoller},\ and\ \citenamefont {Diehl}}]{Budich2015}%
  \BibitemOpen
  \bibfield  {author} {\bibinfo {author} {\bibfnamefont {J.~C.}\ \bibnamefont
  {Budich}}, \bibinfo {author} {\bibfnamefont {P.}~\bibnamefont {Zoller}},\
  and\ \bibinfo {author} {\bibfnamefont {S.}~\bibnamefont {Diehl}},\ }\href
  {https://doi.org/10.1103/PhysRevA.91.042117} {\bibfield  {journal} {\bibinfo
  {journal} {Phys. Rev. A}\ }\textbf {\bibinfo {volume} {91}},\ \bibinfo
  {pages} {042117} (\bibinfo {year} {2015})}\BibitemShut {NoStop}%
\bibitem [{\citenamefont {Iemini}\ \emph {et~al.}(2016)\citenamefont {Iemini},
  \citenamefont {Rossini}, \citenamefont {Fazio}, \citenamefont {Diehl},\ and\
  \citenamefont {Mazza}}]{Iemini2016}%
  \BibitemOpen
  \bibfield  {author} {\bibinfo {author} {\bibfnamefont {F.}~\bibnamefont
  {Iemini}}, \bibinfo {author} {\bibfnamefont {D.}~\bibnamefont {Rossini}},
  \bibinfo {author} {\bibfnamefont {R.}~\bibnamefont {Fazio}}, \bibinfo
  {author} {\bibfnamefont {S.}~\bibnamefont {Diehl}},\ and\ \bibinfo {author}
  {\bibfnamefont {L.}~\bibnamefont {Mazza}},\ }\href
  {https://doi.org/10.1103/PhysRevB.93.115113} {\bibfield  {journal} {\bibinfo
  {journal} {Phys. Rev. B}\ }\textbf {\bibinfo {volume} {93}},\ \bibinfo
  {pages} {115113} (\bibinfo {year} {2016})}\BibitemShut {NoStop}%
\bibitem [{\citenamefont {Linzner}\ \emph {et~al.}(2016)\citenamefont
  {Linzner}, \citenamefont {Wawer}, \citenamefont {Grusdt},\ and\ \citenamefont
  {Fleischhauer}}]{Linzner2016}%
  \BibitemOpen
  \bibfield  {author} {\bibinfo {author} {\bibfnamefont {D.}~\bibnamefont
  {Linzner}}, \bibinfo {author} {\bibfnamefont {L.}~\bibnamefont {Wawer}},
  \bibinfo {author} {\bibfnamefont {F.}~\bibnamefont {Grusdt}},\ and\ \bibinfo
  {author} {\bibfnamefont {M.}~\bibnamefont {Fleischhauer}},\ }\href
  {https://doi.org/10.1103/PhysRevB.94.201105} {\bibfield  {journal} {\bibinfo
  {journal} {Phys. Rev. B}\ }\textbf {\bibinfo {volume} {94}},\ \bibinfo
  {pages} {201105} (\bibinfo {year} {2016})}\BibitemShut {NoStop}%
\bibitem [{\citenamefont {Rivas}\ and\ \citenamefont
  {Martin-Delgado}(2017)}]{Rivas2017}%
  \BibitemOpen
  \bibfield  {author} {\bibinfo {author} {\bibfnamefont {{\'{A}}.}~\bibnamefont
  {Rivas}}\ and\ \bibinfo {author} {\bibfnamefont {M.~A.}\ \bibnamefont
  {Martin-Delgado}},\ }\href {https://doi.org/10.1038/s41598-017-06722-x}
  {\bibfield  {journal} {\bibinfo  {journal} {Sci. Rep.}\ }\textbf {\bibinfo
  {volume} {7}},\ \bibinfo {pages} {6350} (\bibinfo {year} {2017})}\BibitemShut
  {NoStop}%
\bibitem [{\citenamefont {Kawabata}\ \emph {et~al.}(2019)\citenamefont
  {Kawabata}, \citenamefont {Shiozaki}, \citenamefont {Ueda},\ and\
  \citenamefont {Sato}}]{Kawabata2019}%
  \BibitemOpen
  \bibfield  {author} {\bibinfo {author} {\bibfnamefont {K.}~\bibnamefont
  {Kawabata}}, \bibinfo {author} {\bibfnamefont {K.}~\bibnamefont {Shiozaki}},
  \bibinfo {author} {\bibfnamefont {M.}~\bibnamefont {Ueda}},\ and\ \bibinfo
  {author} {\bibfnamefont {M.}~\bibnamefont {Sato}},\ }\href
  {https://doi.org/10.1103/PhysRevX.9.041015} {\bibfield  {journal} {\bibinfo
  {journal} {Phys. Rev. X}\ }\textbf {\bibinfo {volume} {9}},\ \bibinfo {pages}
  {041015} (\bibinfo {year} {2019})}\BibitemShut {NoStop}%
\bibitem [{\citenamefont {Song}\ \emph {et~al.}(2019)\citenamefont {Song},
  \citenamefont {Yao},\ and\ \citenamefont {Wang}}]{Song2019}%
  \BibitemOpen
  \bibfield  {author} {\bibinfo {author} {\bibfnamefont {F.}~\bibnamefont
  {Song}}, \bibinfo {author} {\bibfnamefont {S.}~\bibnamefont {Yao}},\ and\
  \bibinfo {author} {\bibfnamefont {Z.}~\bibnamefont {Wang}},\ }\href
  {https://doi.org/10.1103/PhysRevLett.123.170401} {\bibfield  {journal}
  {\bibinfo  {journal} {Phys. Rev. Lett.}\ }\textbf {\bibinfo {volume} {123}},\
  \bibinfo {pages} {170401} (\bibinfo {year} {2019})}\BibitemShut {NoStop}%
\bibitem [{\citenamefont {Shavit}\ and\ \citenamefont
  {Goldstein}(2020)}]{Shavit2020}%
  \BibitemOpen
  \bibfield  {author} {\bibinfo {author} {\bibfnamefont {G.}~\bibnamefont
  {Shavit}}\ and\ \bibinfo {author} {\bibfnamefont {M.}~\bibnamefont
  {Goldstein}},\ }\href {https://doi.org/10.1103/PhysRevB.101.125412}
  {\bibfield  {journal} {\bibinfo  {journal} {Phys. Rev. B}\ }\textbf {\bibinfo
  {volume} {101}},\ \bibinfo {pages} {125412} (\bibinfo {year}
  {2020})}\BibitemShut {NoStop}%
\bibitem [{\citenamefont {Gau}\ \emph {et~al.}(2020)\citenamefont {Gau},
  \citenamefont {Egger}, \citenamefont {Zazunov},\ and\ \citenamefont
  {Gefen}}]{Gau2020}%
  \BibitemOpen
  \bibfield  {author} {\bibinfo {author} {\bibfnamefont {M.}~\bibnamefont
  {Gau}}, \bibinfo {author} {\bibfnamefont {R.}~\bibnamefont {Egger}}, \bibinfo
  {author} {\bibfnamefont {A.}~\bibnamefont {Zazunov}},\ and\ \bibinfo {author}
  {\bibfnamefont {Y.}~\bibnamefont {Gefen}},\ }\href
  {https://doi.org/10.1103/PhysRevLett.125.147701} {\bibfield  {journal}
  {\bibinfo  {journal} {Phys. Rev. Lett.}\ }\textbf {\bibinfo {volume} {125}},\
  \bibinfo {pages} {147701} (\bibinfo {year} {2020})}\BibitemShut {NoStop}%
\bibitem [{\citenamefont {Lieu}\ \emph {et~al.}(2020)\citenamefont {Lieu},
  \citenamefont {McGinley},\ and\ \citenamefont {Cooper}}]{Lieu2020}%
  \BibitemOpen
  \bibfield  {author} {\bibinfo {author} {\bibfnamefont {S.}~\bibnamefont
  {Lieu}}, \bibinfo {author} {\bibfnamefont {M.}~\bibnamefont {McGinley}},\
  and\ \bibinfo {author} {\bibfnamefont {N.~R.}\ \bibnamefont {Cooper}},\
  }\href {https://doi.org/10.1103/PhysRevLett.124.040401} {\bibfield  {journal}
  {\bibinfo  {journal} {Phys. Rev. Lett.}\ }\textbf {\bibinfo {volume} {124}},\
  \bibinfo {pages} {040401} (\bibinfo {year} {2020})}\BibitemShut {NoStop}%
\bibitem [{\citenamefont {McGinley}\ and\ \citenamefont
  {Cooper}(2020)}]{McGinley2020}%
  \BibitemOpen
  \bibfield  {author} {\bibinfo {author} {\bibfnamefont {M.}~\bibnamefont
  {McGinley}}\ and\ \bibinfo {author} {\bibfnamefont {N.~R.}\ \bibnamefont
  {Cooper}},\ }\href {https://doi.org/10.1038/s41567-020-0956-z} {\bibfield
  {journal} {\bibinfo  {journal} {Nature Phys.}\ }\textbf {\bibinfo {volume}
  {16}},\ \bibinfo {pages} {1181} (\bibinfo {year} {2020})}\BibitemShut
  {NoStop}%
\bibitem [{\citenamefont {Flynn}\ \emph {et~al.}(2021)\citenamefont {Flynn},
  \citenamefont {Cobanera},\ and\ \citenamefont {Viola}}]{Flynn2021}%
  \BibitemOpen
  \bibfield  {author} {\bibinfo {author} {\bibfnamefont {V.~P.}\ \bibnamefont
  {Flynn}}, \bibinfo {author} {\bibfnamefont {E.}~\bibnamefont {Cobanera}},\
  and\ \bibinfo {author} {\bibfnamefont {L.}~\bibnamefont {Viola}},\ }\href
  {https://doi.org/10.1103/PhysRevLett.127.245701} {\bibfield  {journal}
  {\bibinfo  {journal} {Phys. Rev. Lett.}\ }\textbf {\bibinfo {volume} {127}},\
  \bibinfo {pages} {245701} (\bibinfo {year} {2021})}\BibitemShut {NoStop}%
\bibitem [{\citenamefont {Altland}\ \emph {et~al.}(2021)\citenamefont
  {Altland}, \citenamefont {Fleischhauer},\ and\ \citenamefont
  {Diehl}}]{Altland2021}%
  \BibitemOpen
  \bibfield  {author} {\bibinfo {author} {\bibfnamefont {A.}~\bibnamefont
  {Altland}}, \bibinfo {author} {\bibfnamefont {M.}~\bibnamefont
  {Fleischhauer}},\ and\ \bibinfo {author} {\bibfnamefont {S.}~\bibnamefont
  {Diehl}},\ }\href {https://doi.org/10.1103/PhysRevX.11.021037} {\bibfield
  {journal} {\bibinfo  {journal} {Phys. Rev. X}\ }\textbf {\bibinfo {volume}
  {11}},\ \bibinfo {pages} {021037} (\bibinfo {year} {2021})}\BibitemShut
  {NoStop}%
\bibitem [{\citenamefont {Iles-Smith}\ \emph {et~al.}(2014)\citenamefont
  {Iles-Smith}, \citenamefont {Lambert},\ and\ \citenamefont
  {Nazir}}]{IlesSmith2014}%
  \BibitemOpen
  \bibfield  {author} {\bibinfo {author} {\bibfnamefont {J.}~\bibnamefont
  {Iles-Smith}}, \bibinfo {author} {\bibfnamefont {N.}~\bibnamefont
  {Lambert}},\ and\ \bibinfo {author} {\bibfnamefont {A.}~\bibnamefont
  {Nazir}},\ }\href {https://doi.org/10.1103/PhysRevA.90.032114} {\bibfield
  {journal} {\bibinfo  {journal} {Phys. Rev. A}\ }\textbf {\bibinfo {volume}
  {90}},\ \bibinfo {pages} {032114} (\bibinfo {year} {2014})}\BibitemShut
  {NoStop}%
\bibitem [{\citenamefont {Esposito}\ \emph {et~al.}(2015)\citenamefont
  {Esposito}, \citenamefont {Ochoa},\ and\ \citenamefont
  {Galperin}}]{Esposito2015}%
  \BibitemOpen
  \bibfield  {author} {\bibinfo {author} {\bibfnamefont {M.}~\bibnamefont
  {Esposito}}, \bibinfo {author} {\bibfnamefont {M.~A.}\ \bibnamefont
  {Ochoa}},\ and\ \bibinfo {author} {\bibfnamefont {M.}~\bibnamefont
  {Galperin}},\ }\href {https://doi.org/10.1103/PhysRevB.92.235440} {\bibfield
  {journal} {\bibinfo  {journal} {Phys. Rev. B}\ }\textbf {\bibinfo {volume}
  {92}},\ \bibinfo {pages} {235440} (\bibinfo {year} {2015})}\BibitemShut
  {NoStop}%
\bibitem [{\citenamefont {Bruch}\ \emph {et~al.}(2016)\citenamefont {Bruch},
  \citenamefont {Thomas}, \citenamefont {Viola~Kusminskiy}, \citenamefont {von
  Oppen},\ and\ \citenamefont {Nitzan}}]{Bruch2016}%
  \BibitemOpen
  \bibfield  {author} {\bibinfo {author} {\bibfnamefont {A.}~\bibnamefont
  {Bruch}}, \bibinfo {author} {\bibfnamefont {M.}~\bibnamefont {Thomas}},
  \bibinfo {author} {\bibfnamefont {S.}~\bibnamefont {Viola~Kusminskiy}},
  \bibinfo {author} {\bibfnamefont {F.}~\bibnamefont {von Oppen}},\ and\
  \bibinfo {author} {\bibfnamefont {A.}~\bibnamefont {Nitzan}},\ }\href
  {https://doi.org/10.1103/PhysRevB.93.115318} {\bibfield  {journal} {\bibinfo
  {journal} {Phys. Rev. B}\ }\textbf {\bibinfo {volume} {93}},\ \bibinfo
  {pages} {115318} (\bibinfo {year} {2016})}\BibitemShut {NoStop}%
\bibitem [{\citenamefont {Carrega}\ \emph {et~al.}(2016)\citenamefont
  {Carrega}, \citenamefont {Solinas}, \citenamefont {Sassetti},\ and\
  \citenamefont {Weiss}}]{Carrega2016}%
  \BibitemOpen
  \bibfield  {author} {\bibinfo {author} {\bibfnamefont {M.}~\bibnamefont
  {Carrega}}, \bibinfo {author} {\bibfnamefont {P.}~\bibnamefont {Solinas}},
  \bibinfo {author} {\bibfnamefont {M.}~\bibnamefont {Sassetti}},\ and\
  \bibinfo {author} {\bibfnamefont {U.}~\bibnamefont {Weiss}},\ }\href
  {https://doi.org/10.1103/PhysRevLett.116.240403} {\bibfield  {journal}
  {\bibinfo  {journal} {Phys. Rev. Lett.}\ }\textbf {\bibinfo {volume} {116}},\
  \bibinfo {pages} {240403} (\bibinfo {year} {2016})}\BibitemShut {NoStop}%
\bibitem [{\citenamefont {Strasberg}\ \emph {et~al.}(2016)\citenamefont
  {Strasberg}, \citenamefont {Schaller}, \citenamefont {Lambert},\ and\
  \citenamefont {Brandes}}]{Strasberg2016}%
  \BibitemOpen
  \bibfield  {author} {\bibinfo {author} {\bibfnamefont {P.}~\bibnamefont
  {Strasberg}}, \bibinfo {author} {\bibfnamefont {G.}~\bibnamefont {Schaller}},
  \bibinfo {author} {\bibfnamefont {N.}~\bibnamefont {Lambert}},\ and\ \bibinfo
  {author} {\bibfnamefont {T.}~\bibnamefont {Brandes}},\ }\href
  {https://doi.org/10.1088/1367-2630/18/7/073007} {\bibfield  {journal}
  {\bibinfo  {journal} {New J. Phys.}\ }\textbf {\bibinfo {volume} {18}},\
  \bibinfo {pages} {073007} (\bibinfo {year} {2016})}\BibitemShut {NoStop}%
\bibitem [{\citenamefont {Newman}\ \emph {et~al.}(2017)\citenamefont {Newman},
  \citenamefont {Mintert},\ and\ \citenamefont {Nazir}}]{Newman2017}%
  \BibitemOpen
  \bibfield  {author} {\bibinfo {author} {\bibfnamefont {D.}~\bibnamefont
  {Newman}}, \bibinfo {author} {\bibfnamefont {F.}~\bibnamefont {Mintert}},\
  and\ \bibinfo {author} {\bibfnamefont {A.}~\bibnamefont {Nazir}},\ }\href
  {https://doi.org/10.1103/PhysRevE.95.032139} {\bibfield  {journal} {\bibinfo
  {journal} {Phys. Rev. E}\ }\textbf {\bibinfo {volume} {95}},\ \bibinfo
  {pages} {032139} (\bibinfo {year} {2017})}\BibitemShut {NoStop}%
\bibitem [{\citenamefont {Perarnau-Llobet}\ \emph {et~al.}(2018)\citenamefont
  {Perarnau-Llobet}, \citenamefont {Wilming}, \citenamefont {Riera},
  \citenamefont {Gallego},\ and\ \citenamefont {Eisert}}]{PerarnauLlobet2018}%
  \BibitemOpen
  \bibfield  {author} {\bibinfo {author} {\bibfnamefont {M.}~\bibnamefont
  {Perarnau-Llobet}}, \bibinfo {author} {\bibfnamefont {H.}~\bibnamefont
  {Wilming}}, \bibinfo {author} {\bibfnamefont {A.}~\bibnamefont {Riera}},
  \bibinfo {author} {\bibfnamefont {R.}~\bibnamefont {Gallego}},\ and\ \bibinfo
  {author} {\bibfnamefont {J.}~\bibnamefont {Eisert}},\ }\href
  {https://doi.org/10.1103/PhysRevLett.120.120602} {\bibfield  {journal}
  {\bibinfo  {journal} {Phys. Rev. Lett.}\ }\textbf {\bibinfo {volume} {120}},\
  \bibinfo {pages} {120602} (\bibinfo {year} {2018})}\BibitemShut {NoStop}%
\bibitem [{\citenamefont {Miller}\ and\ \citenamefont
  {Anders}(2018)}]{Miller2018}%
  \BibitemOpen
  \bibfield  {author} {\bibinfo {author} {\bibfnamefont {H.~J.~D.}\
  \bibnamefont {Miller}}\ and\ \bibinfo {author} {\bibfnamefont
  {J.}~\bibnamefont {Anders}},\ }\href
  {https://doi.org/10.1038/s41467-018-04536-7} {\bibfield  {journal} {\bibinfo
  {journal} {Nature Commun.}\ }\textbf {\bibinfo {volume} {9}},\ \bibinfo
  {pages} {2203} (\bibinfo {year} {2018})}\BibitemShut {NoStop}%
\bibitem [{\citenamefont {Pancotti}\ \emph {et~al.}(2020)\citenamefont
  {Pancotti}, \citenamefont {Scandi}, \citenamefont {Mitchison},\ and\
  \citenamefont {Perarnau-Llobet}}]{Pancotti2020}%
  \BibitemOpen
  \bibfield  {author} {\bibinfo {author} {\bibfnamefont {N.}~\bibnamefont
  {Pancotti}}, \bibinfo {author} {\bibfnamefont {M.}~\bibnamefont {Scandi}},
  \bibinfo {author} {\bibfnamefont {M.~T.}\ \bibnamefont {Mitchison}},\ and\
  \bibinfo {author} {\bibfnamefont {M.}~\bibnamefont {Perarnau-Llobet}},\
  }\href {https://doi.org/10.1103/PhysRevX.10.031015} {\bibfield  {journal}
  {\bibinfo  {journal} {Phys. Rev. X}\ }\textbf {\bibinfo {volume} {10}},\
  \bibinfo {pages} {031015} (\bibinfo {year} {2020})}\BibitemShut {NoStop}%
\bibitem [{\citenamefont {Rivas}(2020)}]{Rivas2020}%
  \BibitemOpen
  \bibfield  {author} {\bibinfo {author} {\bibfnamefont {A.}~\bibnamefont
  {Rivas}},\ }\href {https://doi.org/10.1103/PhysRevLett.124.160601} {\bibfield
   {journal} {\bibinfo  {journal} {Phys. Rev. Lett.}\ }\textbf {\bibinfo
  {volume} {124}},\ \bibinfo {pages} {160601} (\bibinfo {year}
  {2020})}\BibitemShut {NoStop}%
\bibitem [{\citenamefont {Talkner}\ and\ \citenamefont
  {H\"anggi}(2020)}]{Talkner2020}%
  \BibitemOpen
  \bibfield  {author} {\bibinfo {author} {\bibfnamefont {P.}~\bibnamefont
  {Talkner}}\ and\ \bibinfo {author} {\bibfnamefont {P.}~\bibnamefont
  {H\"anggi}},\ }\href {https://doi.org/10.1103/RevModPhys.92.041002}
  {\bibfield  {journal} {\bibinfo  {journal} {Rev. Mod. Phys.}\ }\textbf
  {\bibinfo {volume} {92}},\ \bibinfo {pages} {041002} (\bibinfo {year}
  {2020})}\BibitemShut {NoStop}%
\bibitem [{\citenamefont {Popovic}\ \emph {et~al.}(2021)\citenamefont
  {Popovic}, \citenamefont {Mitchison}, \citenamefont {Strathearn},
  \citenamefont {Lovett}, \citenamefont {Goold},\ and\ \citenamefont
  {Eastham}}]{Popovic2021}%
  \BibitemOpen
  \bibfield  {author} {\bibinfo {author} {\bibfnamefont {M.}~\bibnamefont
  {Popovic}}, \bibinfo {author} {\bibfnamefont {M.~T.}\ \bibnamefont
  {Mitchison}}, \bibinfo {author} {\bibfnamefont {A.}~\bibnamefont
  {Strathearn}}, \bibinfo {author} {\bibfnamefont {B.~W.}\ \bibnamefont
  {Lovett}}, \bibinfo {author} {\bibfnamefont {J.}~\bibnamefont {Goold}},\ and\
  \bibinfo {author} {\bibfnamefont {P.~R.}\ \bibnamefont {Eastham}},\ }\href
  {https://doi.org/10.1103/PRXQuantum.2.020338} {\bibfield  {journal} {\bibinfo
   {journal} {PRX Quantum}\ }\textbf {\bibinfo {volume} {2}},\ \bibinfo {pages}
  {020338} (\bibinfo {year} {2021})}\BibitemShut {NoStop}%
\bibitem [{\citenamefont {Aidelsburger}\ \emph {et~al.}(2013)\citenamefont
  {Aidelsburger}, \citenamefont {Atala}, \citenamefont {Lohse}, \citenamefont
  {Barreiro}, \citenamefont {Paredes},\ and\ \citenamefont {Bloch}}]{Bloch}%
  \BibitemOpen
  \bibfield  {author} {\bibinfo {author} {\bibfnamefont {M.}~\bibnamefont
  {Aidelsburger}}, \bibinfo {author} {\bibfnamefont {M.}~\bibnamefont {Atala}},
  \bibinfo {author} {\bibfnamefont {M.}~\bibnamefont {Lohse}}, \bibinfo
  {author} {\bibfnamefont {J.~T.}\ \bibnamefont {Barreiro}}, \bibinfo {author}
  {\bibfnamefont {B.}~\bibnamefont {Paredes}},\ and\ \bibinfo {author}
  {\bibfnamefont {I.}~\bibnamefont {Bloch}},\ }\href
  {https://doi.org/10.1103/PhysRevLett.111.185301} {\bibfield  {journal}
  {\bibinfo  {journal} {Phys. Rev. Lett.}\ }\textbf {\bibinfo {volume} {111}},\
  \bibinfo {pages} {185301} (\bibinfo {year} {2013})}\BibitemShut {NoStop}%
\bibitem [{\citenamefont {Rechtsman}\ \emph {et~al.}(2013)\citenamefont
  {Rechtsman}, \citenamefont {Zeuner}, \citenamefont {Plotnik}, \citenamefont
  {Lumer}, \citenamefont {Podolsky}, \citenamefont {Dreisow}, \citenamefont
  {Nolte}, \citenamefont {Segev},\ and\ \citenamefont {Szameit}}]{Szameit}%
  \BibitemOpen
  \bibfield  {author} {\bibinfo {author} {\bibfnamefont {M.~C.}\ \bibnamefont
  {Rechtsman}}, \bibinfo {author} {\bibfnamefont {J.~M.}\ \bibnamefont
  {Zeuner}}, \bibinfo {author} {\bibfnamefont {Y.}~\bibnamefont {Plotnik}},
  \bibinfo {author} {\bibfnamefont {Y.}~\bibnamefont {Lumer}}, \bibinfo
  {author} {\bibfnamefont {D.}~\bibnamefont {Podolsky}}, \bibinfo {author}
  {\bibfnamefont {F.}~\bibnamefont {Dreisow}}, \bibinfo {author} {\bibfnamefont
  {S.}~\bibnamefont {Nolte}}, \bibinfo {author} {\bibfnamefont
  {M.}~\bibnamefont {Segev}},\ and\ \bibinfo {author} {\bibfnamefont
  {A.}~\bibnamefont {Szameit}},\ }\href {https://doi.org/10.1038/nature12066}
  {\bibfield  {journal} {\bibinfo  {journal} {Nature}\ }\textbf {\bibinfo
  {volume} {496}},\ \bibinfo {pages} {196} (\bibinfo {year}
  {2013})}\BibitemShut {NoStop}%
\bibitem [{\citenamefont {Hafezi}\ \emph {et~al.}(2013)\citenamefont {Hafezi},
  \citenamefont {Mittal}, \citenamefont {Fan}, \citenamefont {Migdall},\ and\
  \citenamefont {Taylor}}]{Hafezi2013}%
  \BibitemOpen
  \bibfield  {author} {\bibinfo {author} {\bibfnamefont {M.}~\bibnamefont
  {Hafezi}}, \bibinfo {author} {\bibfnamefont {S.}~\bibnamefont {Mittal}},
  \bibinfo {author} {\bibfnamefont {J.}~\bibnamefont {Fan}}, \bibinfo {author}
  {\bibfnamefont {A.}~\bibnamefont {Migdall}},\ and\ \bibinfo {author}
  {\bibfnamefont {J.~M.}\ \bibnamefont {Taylor}},\ }\href
  {https://doi.org/10.1038/nphoton.2013.274} {\bibfield  {journal} {\bibinfo
  {journal} {Nature Photonics}\ }\textbf {\bibinfo {volume} {7}},\ \bibinfo
  {pages} {1001} (\bibinfo {year} {2013})}\BibitemShut {NoStop}%
\bibitem [{\citenamefont {Jotzu}\ \emph {et~al.}(2014)\citenamefont {Jotzu},
  \citenamefont {Messer}, \citenamefont {Desbuquois}, \citenamefont {Lebrat},
  \citenamefont {Uehlinger}, \citenamefont {Greif},\ and\ \citenamefont
  {Esslinger}}]{EsslingerHaldane}%
  \BibitemOpen
  \bibfield  {author} {\bibinfo {author} {\bibfnamefont {G.}~\bibnamefont
  {Jotzu}}, \bibinfo {author} {\bibfnamefont {M.}~\bibnamefont {Messer}},
  \bibinfo {author} {\bibfnamefont {R.}~\bibnamefont {Desbuquois}}, \bibinfo
  {author} {\bibfnamefont {M.}~\bibnamefont {Lebrat}}, \bibinfo {author}
  {\bibfnamefont {T.}~\bibnamefont {Uehlinger}}, \bibinfo {author}
  {\bibfnamefont {D.}~\bibnamefont {Greif}},\ and\ \bibinfo {author}
  {\bibfnamefont {T.}~\bibnamefont {Esslinger}},\ }\href
  {https://doi.org/10.1038/nature13915} {\bibfield  {journal} {\bibinfo
  {journal} {Nature}\ }\textbf {\bibinfo {volume} {515}},\ \bibinfo {pages}
  {237} (\bibinfo {year} {2014})}\BibitemShut {NoStop}%
\bibitem [{\citenamefont {Aidelsburger}\ \emph {et~al.}(2015)\citenamefont
  {Aidelsburger}, \citenamefont {Lohse}, \citenamefont {Schweizer},
  \citenamefont {Atala}, \citenamefont {Barreiro}, \citenamefont
  {Nascimb{\`{e}}ne}, \citenamefont {Cooper}, \citenamefont {Bloch},\ and\
  \citenamefont {Goldman}}]{Aidelsburger2015}%
  \BibitemOpen
  \bibfield  {author} {\bibinfo {author} {\bibfnamefont {M.}~\bibnamefont
  {Aidelsburger}}, \bibinfo {author} {\bibfnamefont {M.}~\bibnamefont {Lohse}},
  \bibinfo {author} {\bibfnamefont {C.}~\bibnamefont {Schweizer}}, \bibinfo
  {author} {\bibfnamefont {M.}~\bibnamefont {Atala}}, \bibinfo {author}
  {\bibfnamefont {J.~T.}\ \bibnamefont {Barreiro}}, \bibinfo {author}
  {\bibfnamefont {S.}~\bibnamefont {Nascimb{\`{e}}ne}}, \bibinfo {author}
  {\bibfnamefont {N.~R.}\ \bibnamefont {Cooper}}, \bibinfo {author}
  {\bibfnamefont {I.}~\bibnamefont {Bloch}},\ and\ \bibinfo {author}
  {\bibfnamefont {N.}~\bibnamefont {Goldman}},\ }\href
  {https://doi.org/10.1038/nphys3171} {\bibfield  {journal} {\bibinfo
  {journal} {Nat. Phys.}\ }\textbf {\bibinfo {volume} {11}},\ \bibinfo {pages}
  {162} (\bibinfo {year} {2015})}\BibitemShut {NoStop}%
\bibitem [{\citenamefont {Anderson}\ \emph {et~al.}(2016)\citenamefont
  {Anderson}, \citenamefont {Ma}, \citenamefont {Owens}, \citenamefont
  {Schuster},\ and\ \citenamefont {Simon}}]{Schuster}%
  \BibitemOpen
  \bibfield  {author} {\bibinfo {author} {\bibfnamefont {B.~M.}\ \bibnamefont
  {Anderson}}, \bibinfo {author} {\bibfnamefont {R.}~\bibnamefont {Ma}},
  \bibinfo {author} {\bibfnamefont {C.}~\bibnamefont {Owens}}, \bibinfo
  {author} {\bibfnamefont {D.~I.}\ \bibnamefont {Schuster}},\ and\ \bibinfo
  {author} {\bibfnamefont {J.}~\bibnamefont {Simon}},\ }\href
  {https://doi.org/10.1103/PhysRevX.6.041043} {\bibfield  {journal} {\bibinfo
  {journal} {Phys. Rev. X}\ }\textbf {\bibinfo {volume} {6}},\ \bibinfo {pages}
  {041043} (\bibinfo {year} {2016})}\BibitemShut {NoStop}%
\bibitem [{\citenamefont {Khanikaev}\ and\ \citenamefont
  {Shvets}(2017)}]{Khanikaev2017}%
  \BibitemOpen
  \bibfield  {author} {\bibinfo {author} {\bibfnamefont {A.~B.}\ \bibnamefont
  {Khanikaev}}\ and\ \bibinfo {author} {\bibfnamefont {G.}~\bibnamefont
  {Shvets}},\ }\href {https://doi.org/10.1038/s41566-017-0048-5} {\bibfield
  {journal} {\bibinfo  {journal} {Nature Photonics}\ }\textbf {\bibinfo
  {volume} {11}},\ \bibinfo {pages} {763} (\bibinfo {year} {2017})}\BibitemShut
  {NoStop}%
\bibitem [{\citenamefont {Viyuela}\ \emph {et~al.}(2018)\citenamefont
  {Viyuela}, \citenamefont {Rivas}, \citenamefont {Gasparinetti}, \citenamefont
  {Wallraff}, \citenamefont {Filipp},\ and\ \citenamefont
  {Martin-Delgado}}]{Viyuela2018}%
  \BibitemOpen
  \bibfield  {author} {\bibinfo {author} {\bibfnamefont {O.}~\bibnamefont
  {Viyuela}}, \bibinfo {author} {\bibfnamefont {A.}~\bibnamefont {Rivas}},
  \bibinfo {author} {\bibfnamefont {S.}~\bibnamefont {Gasparinetti}}, \bibinfo
  {author} {\bibfnamefont {A.}~\bibnamefont {Wallraff}}, \bibinfo {author}
  {\bibfnamefont {S.}~\bibnamefont {Filipp}},\ and\ \bibinfo {author}
  {\bibfnamefont {M.~A.}\ \bibnamefont {Martin-Delgado}},\ }\href
  {https://doi.org/10.1038/s41534-017-0056-9} {\bibfield  {journal} {\bibinfo
  {journal} {npj Quantum Info.}\ }\textbf {\bibinfo {volume} {4}},\ \bibinfo
  {pages} {10} (\bibinfo {year} {2018})}\BibitemShut {NoStop}%
\bibitem [{\citenamefont {Ozawa}\ \emph {et~al.}(2019)\citenamefont {Ozawa},
  \citenamefont {Price}, \citenamefont {Amo}, \citenamefont {Goldman},
  \citenamefont {Hafezi}, \citenamefont {Lu}, \citenamefont {Rechtsman},
  \citenamefont {Schuster}, \citenamefont {Simon}, \citenamefont {Zilberberg},\
  and\ \citenamefont {Carusotto}}]{TopPhotonReview}%
  \BibitemOpen
  \bibfield  {author} {\bibinfo {author} {\bibfnamefont {T.}~\bibnamefont
  {Ozawa}}, \bibinfo {author} {\bibfnamefont {H.~M.}\ \bibnamefont {Price}},
  \bibinfo {author} {\bibfnamefont {A.}~\bibnamefont {Amo}}, \bibinfo {author}
  {\bibfnamefont {N.}~\bibnamefont {Goldman}}, \bibinfo {author} {\bibfnamefont
  {M.}~\bibnamefont {Hafezi}}, \bibinfo {author} {\bibfnamefont
  {L.}~\bibnamefont {Lu}}, \bibinfo {author} {\bibfnamefont {M.~C.}\
  \bibnamefont {Rechtsman}}, \bibinfo {author} {\bibfnamefont {D.}~\bibnamefont
  {Schuster}}, \bibinfo {author} {\bibfnamefont {J.}~\bibnamefont {Simon}},
  \bibinfo {author} {\bibfnamefont {O.}~\bibnamefont {Zilberberg}},\ and\
  \bibinfo {author} {\bibfnamefont {I.}~\bibnamefont {Carusotto}},\ }\href
  {https://doi.org/10.1103/RevModPhys.91.015006} {\bibfield  {journal}
  {\bibinfo  {journal} {Rev. Mod. Phys.}\ }\textbf {\bibinfo {volume} {91}},\
  \bibinfo {pages} {015006} (\bibinfo {year} {2019})}\BibitemShut {NoStop}%
\bibitem [{\citenamefont {Chalopin}\ \emph {et~al.}(2020)\citenamefont
  {Chalopin}, \citenamefont {Satoor}, \citenamefont {Evrard}, \citenamefont
  {Makhalov}, \citenamefont {Dalibard}, \citenamefont {Lopes},\ and\
  \citenamefont {Nascimbene}}]{Chalopin2020}%
  \BibitemOpen
  \bibfield  {author} {\bibinfo {author} {\bibfnamefont {T.}~\bibnamefont
  {Chalopin}}, \bibinfo {author} {\bibfnamefont {T.}~\bibnamefont {Satoor}},
  \bibinfo {author} {\bibfnamefont {A.}~\bibnamefont {Evrard}}, \bibinfo
  {author} {\bibfnamefont {V.}~\bibnamefont {Makhalov}}, \bibinfo {author}
  {\bibfnamefont {J.}~\bibnamefont {Dalibard}}, \bibinfo {author}
  {\bibfnamefont {R.}~\bibnamefont {Lopes}},\ and\ \bibinfo {author}
  {\bibfnamefont {S.}~\bibnamefont {Nascimbene}},\ }\href
  {https://doi.org/10.1038/s41567-020-0942-5} {\bibfield  {journal} {\bibinfo
  {journal} {Nature Phys.}\ }\textbf {\bibinfo {volume} {16}},\ \bibinfo
  {pages} {1017} (\bibinfo {year} {2020})}\BibitemShut {NoStop}%
\bibitem [{\citenamefont {Viebahn}\ \emph {et~al.}(2021)\citenamefont
  {Viebahn}, \citenamefont {Minguzzi}, \citenamefont {Sandholzer},
  \citenamefont {Walter}, \citenamefont {Sajnani}, \citenamefont {G\"org},\
  and\ \citenamefont {Esslinger}}]{Esslinger_dissipation}%
  \BibitemOpen
  \bibfield  {author} {\bibinfo {author} {\bibfnamefont {K.}~\bibnamefont
  {Viebahn}}, \bibinfo {author} {\bibfnamefont {J.}~\bibnamefont {Minguzzi}},
  \bibinfo {author} {\bibfnamefont {K.}~\bibnamefont {Sandholzer}}, \bibinfo
  {author} {\bibfnamefont {A.-S.}\ \bibnamefont {Walter}}, \bibinfo {author}
  {\bibfnamefont {M.}~\bibnamefont {Sajnani}}, \bibinfo {author} {\bibfnamefont
  {F.}~\bibnamefont {G\"org}},\ and\ \bibinfo {author} {\bibfnamefont
  {T.}~\bibnamefont {Esslinger}},\ }\href
  {https://doi.org/10.1103/PhysRevX.11.011057} {\bibfield  {journal} {\bibinfo
  {journal} {Phys. Rev. X}\ }\textbf {\bibinfo {volume} {11}},\ \bibinfo
  {pages} {011057} (\bibinfo {year} {2021})}\BibitemShut {NoStop}%
\bibitem [{\citenamefont {Ferri}\ \emph {et~al.}(2021)\citenamefont {Ferri},
  \citenamefont {Rosa-Medina}, \citenamefont {Finger}, \citenamefont {Dogra},
  \citenamefont {Soriente}, \citenamefont {Zilberberg}, \citenamefont
  {Donner},\ and\ \citenamefont {Esslinger}}]{Ferri2021}%
  \BibitemOpen
  \bibfield  {author} {\bibinfo {author} {\bibfnamefont {F.}~\bibnamefont
  {Ferri}}, \bibinfo {author} {\bibfnamefont {R.}~\bibnamefont {Rosa-Medina}},
  \bibinfo {author} {\bibfnamefont {F.}~\bibnamefont {Finger}}, \bibinfo
  {author} {\bibfnamefont {N.}~\bibnamefont {Dogra}}, \bibinfo {author}
  {\bibfnamefont {M.}~\bibnamefont {Soriente}}, \bibinfo {author}
  {\bibfnamefont {O.}~\bibnamefont {Zilberberg}}, \bibinfo {author}
  {\bibfnamefont {T.}~\bibnamefont {Donner}},\ and\ \bibinfo {author}
  {\bibfnamefont {T.}~\bibnamefont {Esslinger}},\ }\href
  {https://doi.org/10.1103/PhysRevX.11.041046} {\bibfield  {journal} {\bibinfo
  {journal} {Phys. Rev. X}\ }\textbf {\bibinfo {volume} {11}},\ \bibinfo
  {pages} {041046} (\bibinfo {year} {2021})}\BibitemShut {NoStop}%
\bibitem [{SM()}]{SM}%
  \BibitemOpen
  \href@noop {} {}\bibinfo {note} {See the Supplemental Material for the
  derivation of the exact nonequilibrium steady state, the definitions of the
  currents, a discussion of particle-hole symmetry, a detailed analysis of the
  weak-coupling limit and a semiclassical picture for the edge circulation.
  This includes citations to Refs.~\cite{Chiu2016, Plenio1998,
  MartinezAlvarez2018, Price2012}.}\BibitemShut {Stop}%
\bibitem [{\citenamefont {Dhar}\ \emph {et~al.}(2012)\citenamefont {Dhar},
  \citenamefont {Saito},\ and\ \citenamefont {H\"anggi}}]{Dhar2012}%
  \BibitemOpen
  \bibfield  {author} {\bibinfo {author} {\bibfnamefont {A.}~\bibnamefont
  {Dhar}}, \bibinfo {author} {\bibfnamefont {K.}~\bibnamefont {Saito}},\ and\
  \bibinfo {author} {\bibfnamefont {P.}~\bibnamefont {H\"anggi}},\ }\href
  {https://doi.org/10.1103/PhysRevE.85.011126} {\bibfield  {journal} {\bibinfo
  {journal} {Phys. Rev. E}\ }\textbf {\bibinfo {volume} {85}},\ \bibinfo
  {pages} {011126} (\bibinfo {year} {2012})}\BibitemShut {NoStop}%
\bibitem [{\citenamefont {Ryndyk}(2016)}]{Ryndyk2016}%
  \BibitemOpen
  \bibfield  {author} {\bibinfo {author} {\bibfnamefont {D.}~\bibnamefont
  {Ryndyk}},\ }\href {https://doi.org/10.1007/978-3-319-24088-6} {\emph
  {\bibinfo {title} {Theory of Quantum Transport at Nanoscale}}}\ (\bibinfo
  {publisher} {Springer},\ \bibinfo {address} {Berlin},\ \bibinfo {year}
  {2016})\BibitemShut {NoStop}%
\bibitem [{\citenamefont {Mitchison}\ and\ \citenamefont
  {Plenio}(2018)}]{Mitchison2018}%
  \BibitemOpen
  \bibfield  {author} {\bibinfo {author} {\bibfnamefont {M.~T.}\ \bibnamefont
  {Mitchison}}\ and\ \bibinfo {author} {\bibfnamefont {M.~B.}\ \bibnamefont
  {Plenio}},\ }\href {https://doi.org/10.1088/1367-2630/aa9f70} {\bibfield
  {journal} {\bibinfo  {journal} {New J. Phys.}\ }\textbf {\bibinfo {volume}
  {20}},\ \bibinfo {pages} {033005} (\bibinfo {year} {2018})}\BibitemShut
  {NoStop}%
\bibitem [{\citenamefont {Haldane}(2004)}]{Haldane2004}%
  \BibitemOpen
  \bibfield  {author} {\bibinfo {author} {\bibfnamefont {F.~D.~M.}\
  \bibnamefont {Haldane}},\ }\href
  {https://doi.org/10.1103/PhysRevLett.93.206602} {\bibfield  {journal}
  {\bibinfo  {journal} {Phys. Rev. Lett.}\ }\textbf {\bibinfo {volume} {93}},\
  \bibinfo {pages} {206602} (\bibinfo {year} {2004})}\BibitemShut {NoStop}%
\bibitem [{\citenamefont {Xiao}\ \emph {et~al.}(2010)\citenamefont {Xiao},
  \citenamefont {Chang},\ and\ \citenamefont {Niu}}]{Xiao2010}%
  \BibitemOpen
  \bibfield  {author} {\bibinfo {author} {\bibfnamefont {D.}~\bibnamefont
  {Xiao}}, \bibinfo {author} {\bibfnamefont {M.-C.}\ \bibnamefont {Chang}},\
  and\ \bibinfo {author} {\bibfnamefont {Q.}~\bibnamefont {Niu}},\ }\href
  {https://doi.org/10.1103/RevModPhys.82.1959} {\bibfield  {journal} {\bibinfo
  {journal} {Rev. Mod. Phys.}\ }\textbf {\bibinfo {volume} {82}},\ \bibinfo
  {pages} {1959} (\bibinfo {year} {2010})}\BibitemShut {NoStop}%
\bibitem [{\citenamefont {Xing}\ \emph {et~al.}(2020)\citenamefont {Xing},
  \citenamefont {Xu}, \citenamefont {Balachandran},\ and\ \citenamefont
  {Poletti}}]{Xing2020}%
  \BibitemOpen
  \bibfield  {author} {\bibinfo {author} {\bibfnamefont {B.}~\bibnamefont
  {Xing}}, \bibinfo {author} {\bibfnamefont {X.}~\bibnamefont {Xu}}, \bibinfo
  {author} {\bibfnamefont {V.}~\bibnamefont {Balachandran}},\ and\ \bibinfo
  {author} {\bibfnamefont {D.}~\bibnamefont {Poletti}},\ }\href
  {https://doi.org/10.1103/PhysRevB.102.245433} {\bibfield  {journal} {\bibinfo
   {journal} {Phys. Rev. B}\ }\textbf {\bibinfo {volume} {102}},\ \bibinfo
  {pages} {245433} (\bibinfo {year} {2020})}\BibitemShut {NoStop}%
\bibitem [{\citenamefont {Chiu}\ \emph {et~al.}(2016)\citenamefont {Chiu},
  \citenamefont {Teo}, \citenamefont {Schnyder},\ and\ \citenamefont
  {Ryu}}]{Chiu2016}%
  \BibitemOpen
  \bibfield  {author} {\bibinfo {author} {\bibfnamefont {C.-K.}\ \bibnamefont
  {Chiu}}, \bibinfo {author} {\bibfnamefont {J.~C.~Y.}\ \bibnamefont {Teo}},
  \bibinfo {author} {\bibfnamefont {A.~P.}\ \bibnamefont {Schnyder}},\ and\
  \bibinfo {author} {\bibfnamefont {S.}~\bibnamefont {Ryu}},\ }\href
  {https://doi.org/10.1103/RevModPhys.88.035005} {\bibfield  {journal}
  {\bibinfo  {journal} {Rev. Mod. Phys.}\ }\textbf {\bibinfo {volume} {88}},\
  \bibinfo {pages} {035005} (\bibinfo {year} {2016})}\BibitemShut {NoStop}%
\bibitem [{\citenamefont {Plenio}\ and\ \citenamefont
  {Knight}(1998)}]{Plenio1998}%
  \BibitemOpen
  \bibfield  {author} {\bibinfo {author} {\bibfnamefont {M.~B.}\ \bibnamefont
  {Plenio}}\ and\ \bibinfo {author} {\bibfnamefont {P.~L.}\ \bibnamefont
  {Knight}},\ }\href {https://doi.org/10.1103/RevModPhys.70.101} {\bibfield
  {journal} {\bibinfo  {journal} {Rev. Mod. Phys.}\ }\textbf {\bibinfo {volume}
  {70}},\ \bibinfo {pages} {101} (\bibinfo {year} {1998})}\BibitemShut
  {NoStop}%
\bibitem [{\citenamefont {{Martinez Alvarez}}\ \emph
  {et~al.}(2018)\citenamefont {{Martinez Alvarez}}, \citenamefont {{Barrios
  Vargas}}, \citenamefont {Berdakin},\ and\ \citenamefont {{Foa
  Torres}}}]{MartinezAlvarez2018}%
  \BibitemOpen
  \bibfield  {author} {\bibinfo {author} {\bibfnamefont {V.~M.}\ \bibnamefont
  {{Martinez Alvarez}}}, \bibinfo {author} {\bibfnamefont {J.~E.}\ \bibnamefont
  {{Barrios Vargas}}}, \bibinfo {author} {\bibfnamefont {M.}~\bibnamefont
  {Berdakin}},\ and\ \bibinfo {author} {\bibfnamefont {L.~E.}\ \bibnamefont
  {{Foa Torres}}},\ }\href {https://doi.org/10.1140/epjst/e2018-800091-5}
  {\bibfield  {journal} {\bibinfo  {journal} {Eur. Phys. J. Spec. Top.}\
  }\textbf {\bibinfo {volume} {227}},\ \bibinfo {pages} {1295} (\bibinfo {year}
  {2018})}\BibitemShut {NoStop}%
\bibitem [{\citenamefont {Price}\ and\ \citenamefont
  {Cooper}(2012)}]{Price2012}%
  \BibitemOpen
  \bibfield  {author} {\bibinfo {author} {\bibfnamefont {H.~M.}\ \bibnamefont
  {Price}}\ and\ \bibinfo {author} {\bibfnamefont {N.~R.}\ \bibnamefont
  {Cooper}},\ }\href {https://doi.org/10.1103/PhysRevA.85.033620} {\bibfield
  {journal} {\bibinfo  {journal} {Phys. Rev. A}\ }\textbf {\bibinfo {volume}
  {85}},\ \bibinfo {pages} {033620} (\bibinfo {year} {2012})}\BibitemShut
  {NoStop}%
\end{thebibliography}%
	
	\appendix

\setcounter{secnumdepth}{2}
\setcounter{equation}{0}
\setcounter{figure}{0}
% Add S in front of figure/equation labels
\renewcommand{\theequation}{S\arabic{equation}}
\renewcommand{\thefigure}{S\arabic{figure}}

\section*{\sc \bf Supplemental Material}
\subsection{Steady-state solution}
\label{app:solution}

In this section we briefly detail the solution for the non-equilibrium steady state using the quantum Langevin formalism. We consider a general quadratic Hamiltonian of the form $\hat{H}_{\rm tot} = \hat{H}_S + \hat{H}_{SB} + \hat{H}_B$, with 
\begin{align}\label{H_S}
	\hat{H}_S  & = \sum_{j,k} H_{jk} \adag_j\a_k, \\
	\label{H_SB}
	\hat{H}_{SB} & = \sum_{j\in \partial S} \sum_q g_{qj} \left (\adag_j \b_{qj} + \bdag_{qj} \a_j \right ), \\
	\label{H_B}
	\hat{H}_B & =  \sum_{j\in \partial S} \sum_q \Omega_{qj} \bdag_{qj} \b_{qj}.
\end{align}
In Eq.~\eqref{H_S}, $\a^\dagger_j$ creates a boson or fermion on site $j$ of the system and $H_{jk}$ are the elements of a hermitian single-particle Hamiltonian matrix, $\mathbf{H}$. Similarly, the canonical operators $\bdag_{qj}$ create a particle in mode $q$ of the reservoir connected to site $j$, with corresponding frequency $\Omega_{qj}$ and coupling $g_{qj}$ to a site $j$ lying on the boundary of the system, $\partial S$.

Following the standard procedure, we formally solve the equations of motion for $\b_{qj}$ in the Heisenberg picture and substitute the result into the Heisenberg equation for $\a_j$, obtaining
\begin{equation}\label{QLE_general}
	\ii \partial_t\hat{a}_j(t) =  \sum_k H_{jk} \a_k(t) + \int_{t_0}^\infty\dd t'\, \chi_j(t-t') \a_j(t') + \hat{\xi}_j(t).
\end{equation}
Above, we have introduced the retarded memory kernel $\chi_j(t)$ and the noise operator $\hat{\xi}_j(t)$, which for $j\in \partial S$ are given by
\begin{align}
	\label{noise_op}
	\hat{\xi}_j(t) & = \sum_q g_{qj} \ee^{-\ii \Omega_{qj}(t-t_0)} \b_{qj}(t_0), \\
	\label{memory_kernel}
	\chi_j(t) & = -\ii \theta(t)\sum_q g^2_{qj} \ee^{-\ii \Omega_{qj}t},
\end{align}
with $\theta(t)$ the unit step function, while for $j\notin \partial S$ we define $	\hat{\xi}_j(t) = 0 = \chi_j(t) $. Note that the memory kernel is nothing but the retarded Green function of the noise operator, since $\theta(t-t')\langle [\hat{\xi}_j(t),\hat{\xi}^\dagger_k(t')]_\pm\rangle = \ii  \delta_{jk}\chi_j(t-t')$, where the minus (plus) sign indicates the (anti-)commutator for bosons (fermions). 

Since we are interested in the steady state, we take the limit $t_0\to -\infty$ in Eq.~\eqref{QLE_general} and solve in the Fourier domain. The solution can be expressed in the compact form 
\begin{equation}\label{a_solution_vector}
	\tilde{\mathbf{a}}(\omega) = \mathbf{G}(\omega) \cdot \tilde{\boldsymbol{\xi}}(\omega),
\end{equation}
by defining vectors $\tilde{\mathbf{a}}= (\tilde{a}_1, \tilde{a}_2,\ldots  )^T$ and $\tilde{\bm{\xi}}(\omega) = (\tilde{\xi}_1, \tilde{\xi}_2,\ldots  )^T$ of Fourier-transformed operators and introducing the (matrix-valued) retarded Green function
\begin{equation}\label{G_ret}
	\mathbf{G}(\omega) = \left [ \omega\mathbf{1} - \mathbf{H} -\mathbf{\Sigma}(\omega) \right ]^{-1}.
\end{equation}
Here, $\mathbf{1}$ is the identity matrix and the retarded self-energy matrix $\mathbf{\Sigma}(\omega)$ is diagonal with elements $\Sigma_{jk}(\omega) = \delta_{jk} \tilde{\chi}_j(\omega)$ given by the Fourier transform of the memory kernel. The latter decomposes as 
\begin{equation}\label{memory_kernel_decomp}
	\tilde{\chi}_j(\omega) = \fint \frac{\dd\omega'}{2\pi} \frac{\gamma_j(\omega')}{\omega-\omega'} - \ii \frac{\gamma_j(\omega)}{2},
\end{equation}
where $ \fint$ denotes a principal-value integral and we introduced the spectral density
\begin{equation}\label{spectral_density}
	\gamma_j(\omega) = 2\pi \sum_q g_{qj}^2\delta(\omega-\Omega_{qj}).
\end{equation}

The non-equilibrium steady state of the system depends on the statistical properties of the noise operator via Eq.~\eqref{a_solution_vector}. Assuming that the initial state of the reservoirs (at time $t=t_0$) is thermal and uncorrelated, we find the noise spectrum
\begin{equation}\label{noise_spectrum}
	\langle \tilde{\xi}^\dagger_j(\omega) \tilde{\xi}_k(\omega')\rangle = 2\pi\delta(\omega-\omega')\delta_{jk} \gamma_j(\omega) f_j(\omega),
\end{equation}
where $f_j(\omega) = (\ee^{\beta_j(\omega-\mu_j)} \pm 1 )^{-1}$ is the reservoir distribution function, characterized by an inverse temperature $\beta_j$ and chemical potential $\mu_j$, and expectation values are taken with respect to the state at time $t=t_0$. It is now straightforward to write down the two-point correlation function of the system in its steady state, $C_{jk}(t) = \langle \adag_k(t) \a_j(0)\rangle$, as
\begin{equation}\label{two_point_function}
	\mathbf{C}(t) = \int\frac{\dd\omega	}{2\pi} \ee^{\ii\omega t} \mathbf{G} \cdot (\mathbf{\Gamma} \circ \mathbf{F}) \cdot \mathbf{G}^\dagger.
\end{equation}
Here we have suppressed frequency arguments and defined diagonal matrices $\mathbf{\Gamma} = \ii (\mathbf{\Sigma} - \mathbf{\Sigma}^\dagger)$ and $F_{jk} = \delta_{jk} f_j$, while $\circ$ denotes the elementwise (Hadamard) product. The steady-state correlation matrix $\mathbf{C} = \mathbf{C}(0)$ is found simply by setting $t=0$ in Eq.~\eqref{two_point_function}.

The setup considered in the main text features hot baths coupled to sites on the boundary at $x=1$ and cold baths coupled to the other boundary at $x=L_X$. Moreover, the wide-band limit approximation is made, which amounts to replacing the spectral densities $\gamma_j(\omega)$ by frequency-independent constants. This approximation is valid so long as $\gamma_j(\omega)$ does not vary significantly over the frequency range within which the integrand of Eq.~\eqref{two_point_function} differs appreciably from zero.  Assuming symmetric coupling, $\gamma_j = \gamma$, one then recovers Eq.~(4). In this case, $\mathbf{\Gamma} = \mathbf{\Gamma}_h + \mathbf{\Gamma}_c$, where the self-energy $\mathbf{\Gamma}_h$ ($\mathbf{\Gamma}_c$) of the hot (cold) bath is given by a diagonal matrix, with $[\mathbf{\Gamma}_{h,c}]_{jk} = \gamma \delta_{jk}$ whenever $j$ corresponds to a boundary site with $x=1$ ($x=L_X$) and $[\mathbf{\Gamma}_{h,c}]_{jk} = 0$  otherwise. Likewise, we have $f_j = \bar{n}_h$ ($f_j = \bar{n}_c$) for $j$ corresponding to $x=1$ ($x=L_X$).

\subsection{Definition of currents}
\label{app:currents}

The current densities are defined via the continuity equation for the particle density, $\hat{n}_{x,y} = {\bf\a}^\dagger_{x,y}
\cdot {\bf \a}_{x,y}$, as
\begin{align}\label{continuity_equations}
	\ii [\hat{H}_X, \hat{n}_{x,y}] & = \hat{J}^X_{x-1,y} - \hat{J}^X_{x,y} \notag \\
	\ii [\hat{H}_Y, \hat{n}_{x,y}] 	& =  \hat{J}^Y_{x,y-1} - \hat{J}^Y_{x,y}.
\end{align}
Here $\hat{J}^X_{x,y}$ denotes the current density operator describing particle flow from site $(x,y)$ to its nearest neighbour in the $x$ direction, i.e.~$(x,y) \to (x+1,y)$. Similarly, $\hat{J}^Y_{x,y}$ is the corresponding current in the $y$ direction, $(x,y) \to (x,y+1)$. Using the canonical (anti-)commutation relations, it is straightforward to derive
\begin{align}
	\label{current_op_X}
	\hat{J}^X_{x,y} & = \frac{t_X}{2\ii } {\bf \a}^\dagger_{x+1,y}\cdot (\sigma_z+\ii\sigma_y)\cdot {\bf \a}_{x,y} + {\rm h.c.},\\ 
	\label{current_op_Y}
	\hat{J}^Y_{x,y} & = \frac{t_Y}{2\ii } {\bf \a}^\dagger_{x,y+1}\cdot (\sigma_z+\ii\sigma_x)\cdot {\bf \a}_{x,y} + {\rm h.c.}
\end{align}
The expectation values of these one-body observables can then be read off directly from the elements of the steady-state correlation matrix.

The total particle current flowing, say, into the cold bath is represented by the observable $\hat{J}_{\rm tot} = \ii[\hat{H}_{\rm tot}, \hat{N}_c]$, where $\hat{N}_c = \sum_{j\in c}\sum_q \hat{b}_{qj}^\dag \hat{b}_{qj}$ is the particle-number operator of the cold bath. Following a similar calculation to Sec.~\ref{app:solution}, we find the standard Landauer formula for the mean steady-state current~\cite{Ryndyk2016}
\begin{equation}\label{J_tot}
J_{\rm tot} = \int\frac{\dd	\omega}{2\pi} \,\mathcal{T}(\omega) \left [\bar{n}_h(\omega) - \bar{n}_c(\omega)\right ],
\end{equation}
where $\mathcal{T}(\omega) = \Tr\left [ \mathbf{\Gamma}_c\cdot \mathbf{G} \cdot\mathbf{\Gamma}_h\cdot \mathbf{G}^\dagger\right ]$ is the transmission function. Particle conservation implies that $\sum_y J^X_{x,y} = J_{\rm tot}$ in the NESS. 

\subsection{Particle-hole symmetry}
\label{app:PH_symmetry}

In the fermionic setting, the system obeys a generalized particle-hole (GPH) symmetry \cite{Chiu2016} when $\mu=\omega_0$. It is simplest to first analyse this situation assuming that $\omega_0=0$.  In this case,  the fermionic Hamiltonian [Eqs.~(1)--(3)] is invariant under the combined flavor swap and particle hole transformation,  $\hat{\Upsilon}\mathbf{\hat{a}}_{x,y}^\dagger\hat{\Upsilon}^{\dagger} = (\sigma_x \mathbf{\hat{a}}_{x,y})^{\rm T}$ or, explicitly,  $\hat{\Upsilon}(\hat{a}^\dagger_{x,y,\uparrow}, \hat{a}^\dagger_{x,y,\downarrow})\hat{\Upsilon}^{\dagger} = (\hat{a}_{x,y,\downarrow}, \hat{a}_{x,y,\uparrow})$. Meanwhile, the current operators defined in Eqs.~\eqref{current_op_X} and~\eqref{current_op_Y} are odd under this transformation, i.e.~$\hat{\Upsilon}\hat{J}^{X,Y}_{x,y} \hat{\Upsilon}^{\dagger} = -\hat{J}^{X,Y}_{x,y}$. It follows that any state $\hat{\rho}$ that is GPH-invariant, satisfying $[\hat{\rho},\hat{\Upsilon}] = 0$, has vanishing currents. Indeed, schematically we have $\braket{\hat{J}} = \Tr[\hat{J}\hat{\rho}] = -\Tr[\hat{\Upsilon}\hat{J}\hat{\Upsilon}^{\dagger}\hat{\rho}] = -\braket{\hat{J}}$, where $\hat{J}$ stands for any current operator that is GPH-odd.

In general, the correlation matrix can be written in block form as
\begin{equation}\label{C_block}
	\mathbf{C} = \left ( \begin{matrix}
		\mathbf{C}_{\uparrow \uparrow} & \mathbf{C}_{\uparrow \downarrow}\\		\mathbf{C}_{\downarrow \uparrow} & \mathbf{C}_{\downarrow \downarrow}
	\end{matrix}\right ),
\end{equation}
where $\mathbf{C}_{\uparrow \uparrow}$ is the correlation matrix for the $\uparrow$ flavor, $\mathbf{C}_{\uparrow \downarrow} = \mathbf{C}_{\downarrow \uparrow}^\dagger $ describes coherences between the different flavors, etc. The condition for GPH-invariance, $[\hat{\rho},\hat{\Upsilon}] = 0$, implies that the fermionic correlation matrix satisfies
\begin{equation}\label{PH_symmetry_C}
	\mathbf{C} = \mathbf{1} - \mathbf{S}\cdot\mathbf{C}^*\cdot \mathbf{S},
\end{equation}
where $\mathbf{S}$ is a matrix that swaps the flavors, and $\mathbf{C}^*$ is the complex conjugate of $\mathbf{C}$. With the basis order implicit in Eq.~\eqref{C_block}, we have
\begin{equation}\label{S_matrix}
	\mathbf{S} = \left ( \begin{matrix}
		 \mathbf{0} & \mathbf{1}\\	\mathbf{1} & \mathbf{0}
	\end{matrix}\right ),
\end{equation}
where $\mathbf{0}$ is the zero matrix. The GPH symmetry of the many-body Hamiltonian translates to the condition
\begin{equation}\label{PH_single_particle_H}
	\mathbf{H} = -\mathbf{S}\cdot\mathbf{H}^*\cdot\mathbf{S}.
\end{equation}
In turn, this implies the following constraint on the retarded Green function:
\begin{equation}\label{PH_Green_function}
	\mathbf{G}(-\omega) = - \mathbf{S}\cdot\left [\mathbf{G}(\omega)\right ]^*\cdot\mathbf{S}.
\end{equation}
Note that here we assume the self-energy is also GPH-symmetric, i.e.~$\mathbf{\Sigma}(\omega) = -\mathbf{S}\cdot[\mathbf{\Sigma}(-\omega)]^*\cdot\mathbf{S}$. In particular, this applies when the system-bath coupling is identical for both flavor states, so that $[\mathbf{\Sigma},\mathbf{S}] =0$, and when the spectral densities are symmetric, $\gamma_j(\omega) = \gamma_j(-\omega)$.

At the GPH-symmetric point, $\mu=0$, the distribution functions satisfy $\bar{n}_{h,c}(\omega) =  1 - \bar{n}_{h,c}(-\omega)$. Thus, reversing the dummy integration variable as $\omega \to -\omega$ in Eq.~(4), we obtain
\begin{align}
	\label{C_PH_derivation}
\mathbf{C} & = \int\frac{\dd\omega}{2\pi} \, \mathbf{G}(\omega) \cdot \Big [\mathbf{ \Gamma}_h \bar{n}_h(\omega) + \mathbf{\Gamma}_c \bar{n}_c(\omega)\Big]\cdot \mathbf{G}^\dagger(\omega)\notag \\
 & = \int\frac{\dd\omega}{2\pi} \, \mathbf{G}(\omega) \cdot \Big [\mathbf{ \Gamma} - \mathbf{ \Gamma}_h \bar{n}_h(-\omega) - \mathbf{\Gamma}_c \bar{n}_c(-\omega)\Big]\cdot\mathbf{G}^\dagger(\omega)\notag \\
& =  \mathbf{1} - \int\frac{\dd\omega}{2\pi} \, \mathbf{G}(-\omega) \cdot \Big[ \mathbf{ \Gamma}_h \bar{n}_h(\omega) + \mathbf{\Gamma}_c \bar{n}_c(\omega)\Big]\cdot\mathbf{G}^\dagger(-\omega)\notag \\
& =  \mathbf{1} - \mathbf{S}\cdot\mathbf{C}^*\cdot \mathbf{S}.
\end{align}
Above, we have made use of Eq.~\eqref{PH_Green_function} as well as the identity $\mathbf{G}\cdot \mathbf{\Gamma}\cdot \mathbf{G}^\dagger = \ii \left (\textbf{G} - \mathbf{G}^\dagger\right )$ and the property
\begin{equation}\label{commutation_Gret}
\int\frac{\dd\omega}{2\pi }\ii  \left (\textbf{G} - \mathbf{G}^\dagger \right ) = \mathbf{1},
\end{equation}
which follows from the (anti-)commutation relations~\cite{Kamenev2009}. Since the NESS is the unique Gaussian state with correlation matrix given by Eq.~\eqref{C_PH_derivation}, it follows that the state is GPH-symmetric when $\mu= \omega_0 = 0$. This implies that all currents vanish. Consistent with this observation, GPH symmetry implies that the transmission function is even, $\mathcal{T}(\omega) = \mathcal{T}(-\omega)$, and therefore the integral in Eq.~\eqref{J_tot} is odd and equates to zero. 

The same GPH symmetries and associated vanishing currents are obtained when $\mu = \omega_0 \neq 0$. This can be shown simply by shifting the frequency variable $\omega\to \omega +\mu$ in Eq.~\eqref{C_PH_derivation}. The final result then follows from the GPH symmetry of the shifted Hamiltonian $\mathbf{H}' = \mathbf{H}-\omega_0\mathbf{1}$. We have numerically confirmed the GPH symmetry of the NESS in the wide-band limit when $\mu=\omega_0$, both in weak- and strong-coupling regimes.

\subsection{Weak-coupling limit}
\label{app:weak_coupling}

Following Ref.~\cite{Dhar2012}, we show that in the weak-coupling limit the NESS reduces to an energy-diagonal density matrix that coincides with the solution of the secular-Born-Markov master equation~\cite{Breuer2002,RivasHuelga,Rivas2017}. The system Hamiltonian can be written in its spectral representation
\begin{equation}\label{single_particle_H_diag}
	\mathbf{H} = \mathbf{U}\cdot\mathbf{\Omega}\cdot\mathbf{U}^\dagger,	
\end{equation}
where $\mathbf{\Omega} = {\rm diag}\{\omega_\alpha\}$ and $\mathbf{U}$ comprise the eigenvalues and eigenvectors of $\mathbf{H}$. The many-body Hamiltonian in this basis reads as
\begin{equation}\label{many_body_H_diag}
\hat{H} = \sum_{\alpha} \omega_\alpha \cdag_\alpha \c_\alpha, \qquad \c_\alpha = \sum_{j} U^*_{j\alpha}\a_j	.
\end{equation}

Now consider the retarded Green function~\eqref{G_ret} in the wide-band limit, $\mathbf{\Sigma} = -\ii \mathbf{\Gamma}/2 \propto \gamma$. As $\gamma\to 0$, $\mathbf{G}(\omega)$ becomes approximately diagonal in the energy representation. Following identical arguments to Ref.~\cite{Dhar2012}, we obtain the steady-state correlation matrix in the energy eigenbasis as
\begin{equation}\label{diagonal_NESS}
\braket{\cdag_\alpha \c_{\alpha'}} = \delta_{\alpha\alpha'} \frac{s_\alpha \bar{n}_h(\omega_\alpha) + r_\alpha \bar{n}_c(\omega_\alpha)}{s_\alpha + r_\alpha},	
\end{equation}
where $s_\alpha$ and $r_\alpha$ describe the dimensionless coupling strength of eigenmode $\alpha$ to the hot and cold baths, respectively:
\begin{align}
\label{s_alpha}
& s_\alpha = \gamma^{-1}\sum_{j,k}U^*_{j\alpha} [\Gamma_h]_{jk}	U_{k\alpha} = \sum_{y=1}^{L_Y}\sum_{\sigma=\uparrow,\downarrow} |U_{(1,y,\sigma),\alpha}|^2,\\
\label{r_alpha}
& r_\alpha = \gamma^{-1}\sum_{j,k}U^*_{j\alpha} [\Gamma_c]_{jk}	U_{k\alpha} = \sum_{y=1}^{L_Y}\sum_{\sigma=\uparrow,\downarrow} |U_{(L_X,y,\sigma),\alpha}|^2.
\end{align}

We now consider the Lindblad master equation obtained within the Born-Markov and secular approximations. Following the standard derivation~\cite{Breuer2002,RivasHuelga} yields the master equation
\begin{equation}\label{Lindblad}
	\dot{\hat{\rho}} = -\ii [\hat{H}, \hat{\rho}] + \mathcal{L}_h\hat{\rho} + \mathcal{L}_c\hat{\rho},
\end{equation}
with dissipators for the hot and cold baths given respectively by
\begin{align}
	\label{hot_dissipator}
	\mathcal{L}_h = \gamma \sum_\alpha s_\alpha \left ( \bar{n}_h(\omega_\alpha)\mathcal{D}[\cdag_\alpha] + [1\mp \bar{n}_h(\omega_\alpha)]\mathcal{D}[\c_\alpha]  \right ),\\
		\label{cold_dissipator}
	\mathcal{L}_c = \gamma \sum_\alpha r_\alpha \left ( \bar{n}_c(\omega_\alpha)\mathcal{D}[\cdag_\alpha] + [1\mp \bar{n}_c(\omega_\alpha)]\mathcal{D}[\c_\alpha]  \right ),
\end{align}
where $\mathcal{D}[\hat{L}]\bullet  = \hat{L}\bullet \hat{L}^\dagger - \tfrac{1}{2}\{\hat{L}^\dagger \hat{L},\bullet\}$ is a Lindblad superoperator and the plus (minus) sign is for bosons (fermions). It is easy to check that the steady-state solution of Eq.~\eqref{Lindblad}, $\dot{\hat{\rho}}=0$, is a Gaussian state with correlation matrix given exactly by Eq.~\eqref{diagonal_NESS}. 

As is well known, the Lindblad equation~\eqref{Lindblad} can be written as 
\begin{equation}\label{Lindblad_nonHermitian}
\dot{\hat{\rho}} = -\ii \hat{H}_{\rm eff} \hat{\rho} + \ii \hat{\rho} \hat{H}_{\rm eff}^\dagger + \mathcal{J}_h\hat{\rho} + \mathcal{J}_c\hat{\rho},
\end{equation}
where the non-Hermitian Hamiltonian is
\begin{align}\label{H_nonHermitian}
\hat{H}_{\rm eff} & = \hat{H} - \frac{\ii}{2} \sum_{\alpha} \gamma s_\alpha \left[ \bar{n}_h(\omega_\alpha) + (1\mp 2\bar{n}_{h}(\omega_\alpha)) \hat{c}^\dagger_\alpha \hat{c}_\alpha\right ] \notag \\ 
& \qquad - \frac{\ii}{2} \sum_{\alpha} \gamma r_\alpha \left[ \bar{n}_c(\omega_\alpha) + (1\mp 2\bar{n}_{c}(\omega_\alpha)) \hat{c}^\dagger_\alpha \hat{c}_\alpha\right ],
\end{align}
and the superoperators $\mathcal{J}_{c,h}$ represent ``quantum jumps'' due to exchange of particles with the baths~\cite{Plenio1998},
\begin{align}
\label{jump_superoperators}
\mathcal{J}_h\hat{\rho} & = \sum_\alpha \gamma s_\alpha \left [\bar{n}_h(\omega_\alpha) \hat{c}_\alpha^\dagger \hat{\rho}\hat{c}_\alpha + \left (1 + \bar{n}_h(\omega_\alpha)\right ) \hat{c}_\alpha \hat{\rho}\hat{c}_\alpha^\dagger  \right ] ,\\
\mathcal{J}_c\hat{\rho} & = \sum_\alpha \gamma r_\alpha \left [\bar{n}_c(\omega_\alpha) \hat{c}_\alpha^\dagger \hat{\rho}\hat{c}_\alpha + \left (1 + \bar{n}_c(\omega_\alpha)\right ) \hat{c}_\alpha \hat{\rho}\hat{c}_\alpha^\dagger  \right ] .
\end{align}
By including both the non-Hermitian Hamiltonian and the jump terms, we ensure a completely positive and physically consistent description of the full many-body state of the non-equilibrium lattice system, valid at arbitrary times. This is in contrast to non-Hermitian models that neglect the jumps (valid for rare post-selected trajectories of continuously measured systems) or the effective non-Hermitian Hamiltonians appearing in single-particle Green functions, the topological properties of which have recently been of intense interest; e.g., see Ref.~\cite{MartinezAlvarez2018} and references therein.

\begin{figure*}
	\begin{minipage}{0.31\linewidth}
		\includegraphics[width=\linewidth]{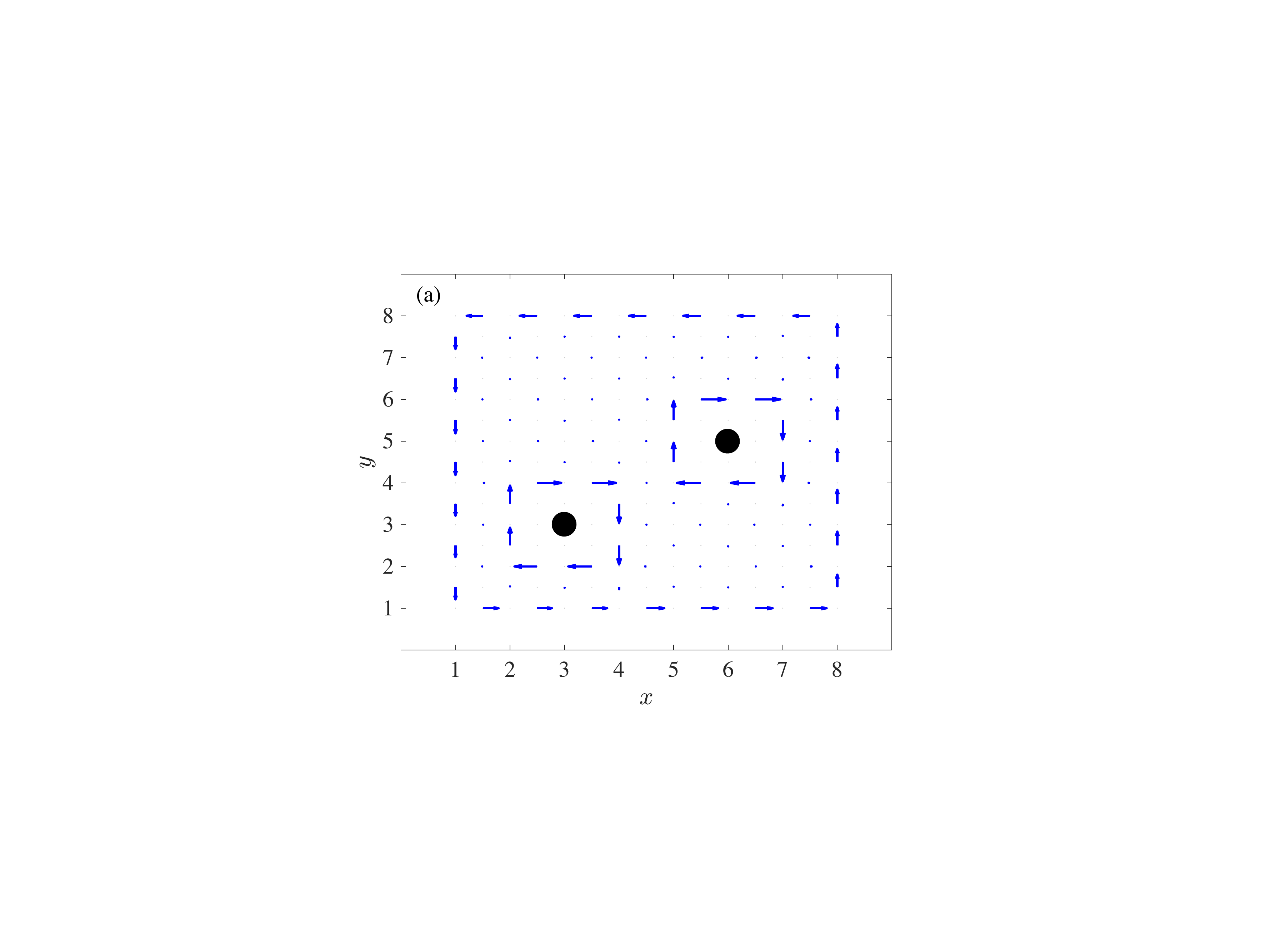}
	\end{minipage}\quad 
	\begin{minipage}{0.31\linewidth}
		\includegraphics[width=\linewidth]{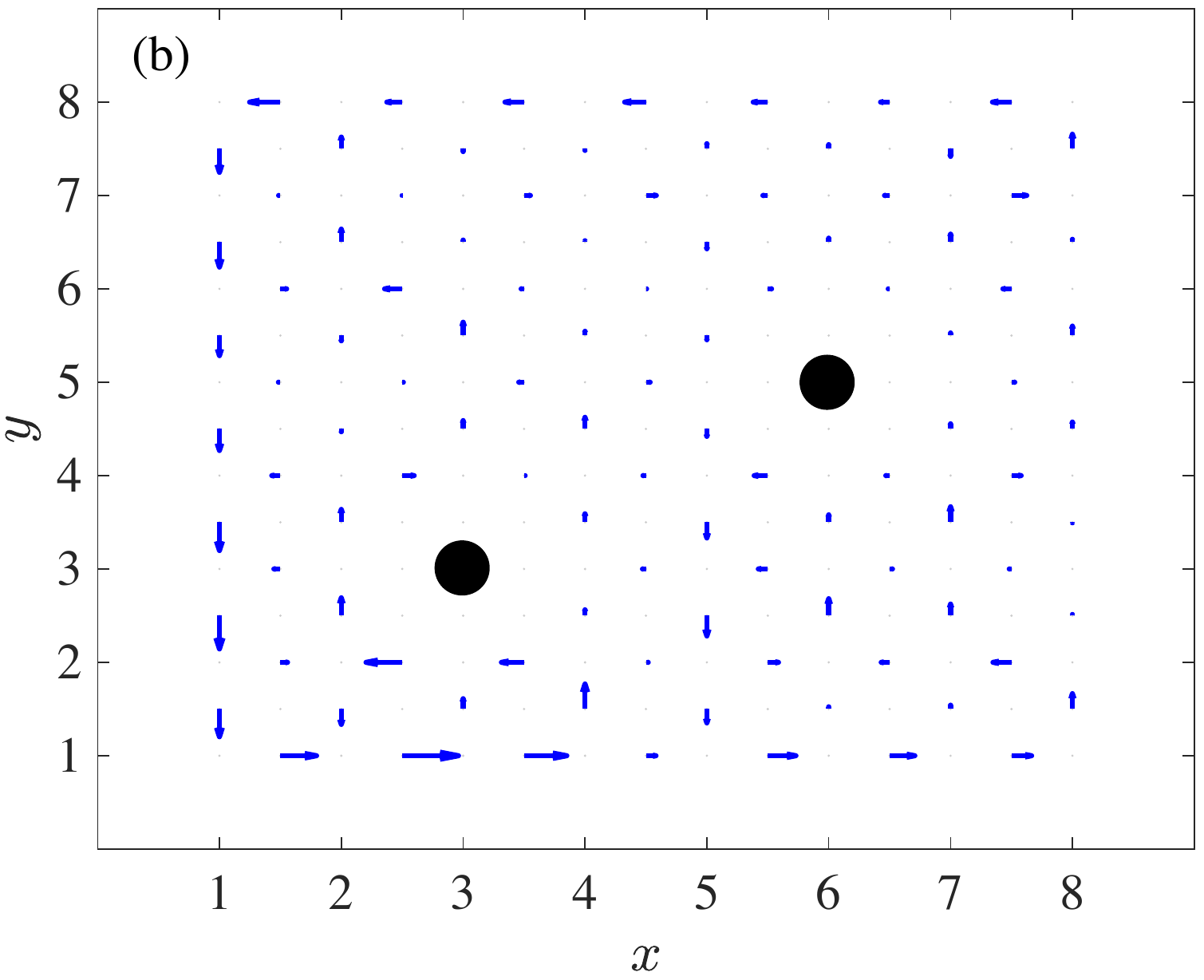}
	\end{minipage}\quad
	\begin{minipage}{0.31\linewidth}
		\includegraphics[width=\linewidth]{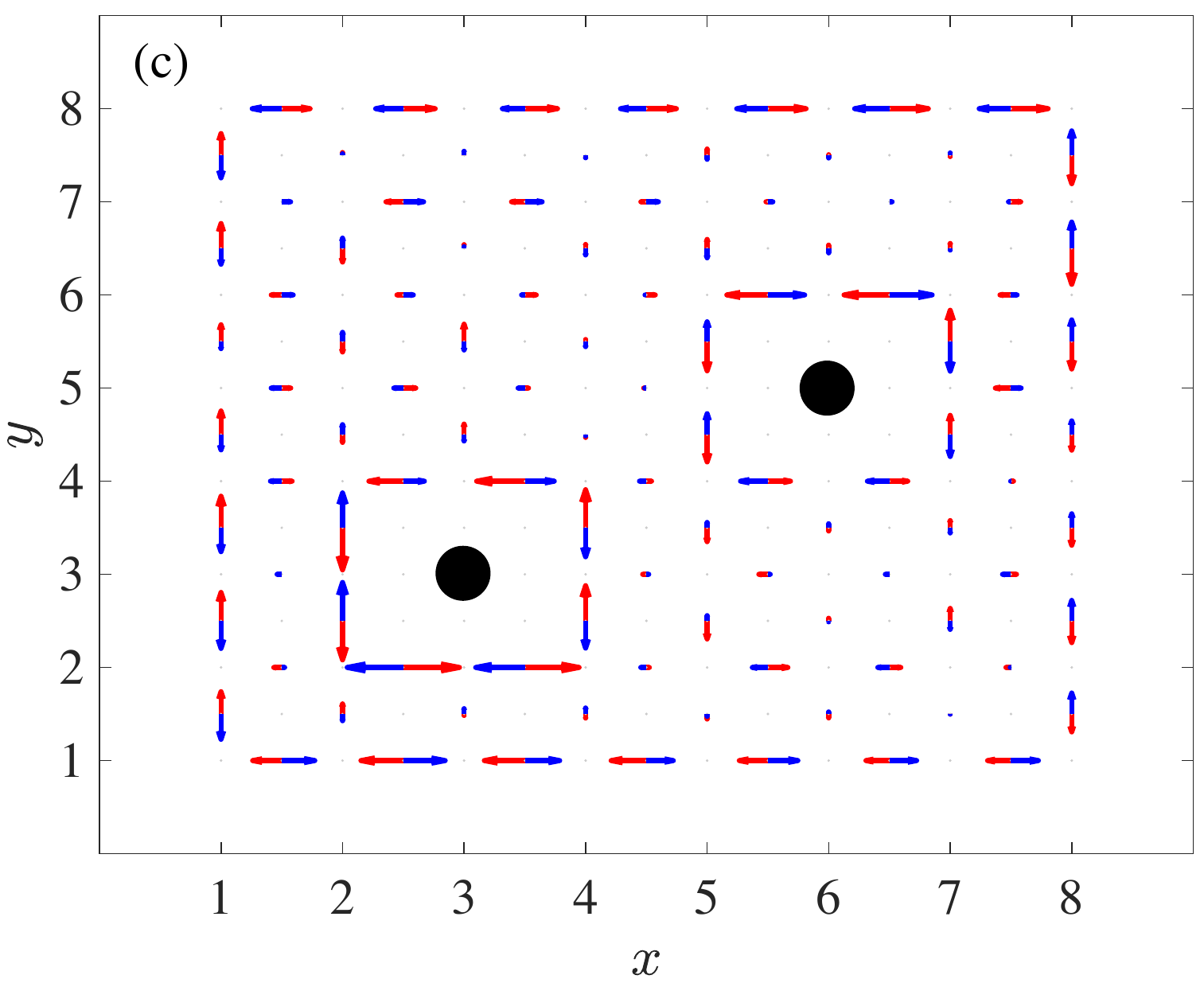}
	\end{minipage}
	\caption{Current patterns for the same asymmetric impurity distribution shown in Fig.~4(c) but with different parameters. (a,b) Fermionic currents in the topologically nontrivial regime, $m=t$, with on-site energy $\omega_0=10t$, chemical potential $\mu = \omega-0.5t$, and (a) low temperatures, $T_h=t$ and $T_c=0.01t$, or (b) higher temperatures, $T_h=100t$ and $T_c=t$. (c) Same as (a) but showing bosons (red) and fermions (blue) in a topologically trivial regime, $m=3t$. \label{fig:fermion_impurity_quiver}}
\end{figure*}

\subsection{Weak-coupling distribution functions}
\label{app:distribution}

\begin{figure}[b]
	\begin{minipage}{0.495\linewidth}
		\includegraphics[width=\linewidth]{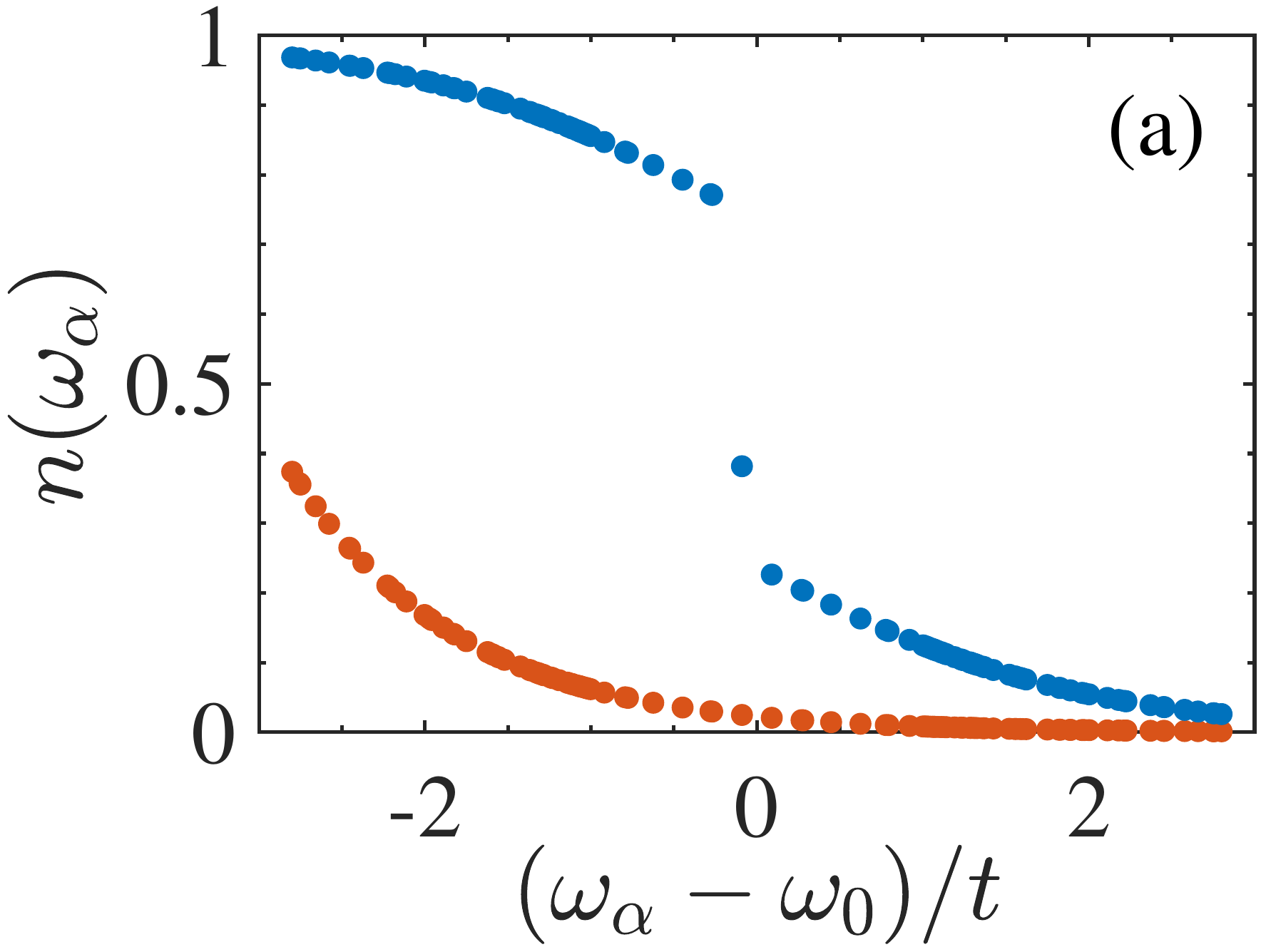}
	\end{minipage}
	\begin{minipage}{0.495\linewidth}
		\includegraphics[width=\linewidth]{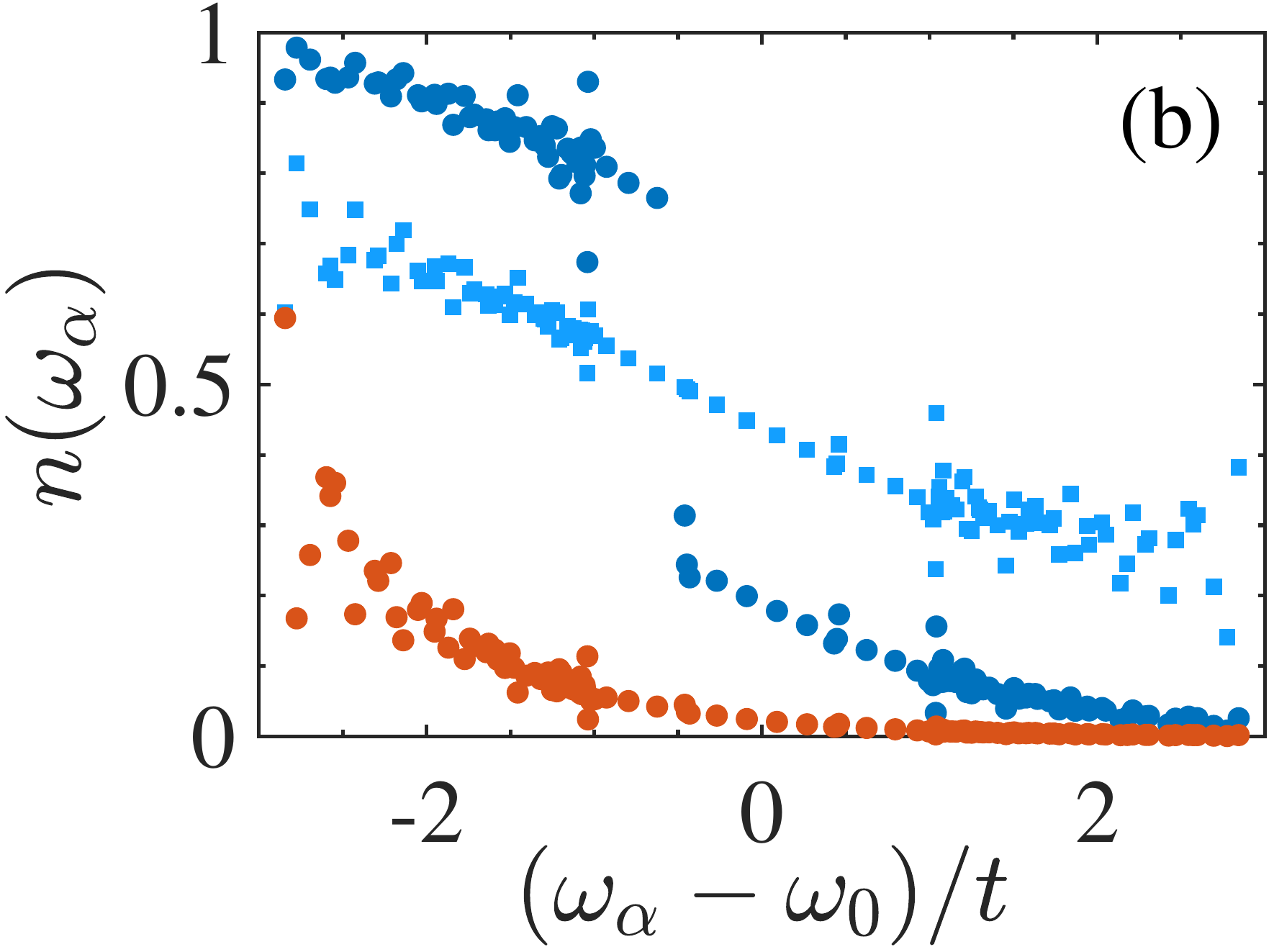}
	\end{minipage}
	\caption{Distribution functions in the NESS for bosons (red) and fermions (blue). (a)~Symmetric impurity configuration shown in Fig.~4(b), with $T_h=t$, $T_c=0.01t$, $\omega_0=10t$ and $\mu=\omega_0-0.1t$. (b)~Asymmetric impurity configuration shown in Fig.~4(c), but with $\mu=\omega_0-0.5t$ as in Fig.~\ref{fig:fermion_impurity_quiver}. Circles show the data for $T_h=t$ and $T_c=0.01t$ [Fig.~\ref{fig:fermion_impurity_quiver}(a)], while light blue squares show the result for fermions at high temperature, $T_h=100t$ and $T_c = t$ [Fig.~\ref{fig:fermion_impurity_quiver}(b)]. The values of $n(\omega_\alpha)$ for bosons are multiplied by $10^3$ to be visible on the same axes. \label{fig:distributions}}
\end{figure}

We now investigate in more detail the form of the weak-coupling distribution function, $n(\omega_\alpha) = \braket{\hat{c}_\alpha^\dagger\hat{c}_\alpha}$, defined by Eq.~\eqref{diagonal_NESS}.  When the reservoirs are at the same temperature, $\bar{n}_h=\bar{n}_c = \bar{n}$, we obtain simply $ n(\omega_\alpha) = \bar{n}(\omega_\alpha)$. This is a smooth, monotonically decreasing function of frequency that is independent of the mode-specific couplings $s_\alpha$ and $r_\alpha$.  As a consequence, the equilibrium currents---as seen, for example, in Fig.~2(c)---are unaffected by the addition of defects in any configuration~\cite{Rivas2017}.

Out of equilibrium, an analogous result is obtained in the clean QWZ model described by Eqs.~(1)--(3) due to the symmetries $\hat{\Pi}\hat{\Theta}\hat{\Sigma}_y$ and $\hat{\Pi}\hat{R}_\pi$. In particular, the symmetry $\hat{\Pi}\hat{\Theta}\hat{\Sigma}_y$ implies that $|U_{(1,y,\sigma),\alpha}|^2 = |U_{(L_X,y,\sigma),\alpha}|^2 $ and therefore $s_\alpha = r_\alpha$ identically. Similarly, the symmetry under $\hat{\Pi}\hat{R}_\pi$ means that $|U_{(1,y,\sigma),\alpha}|^2 = |U_{(L_X,L_Y+1-y,\sigma),\alpha}|^2$, and again $s_\alpha = r_\alpha$ upon summing over $y$ in Eqs.~\eqref{s_alpha} and~\eqref{r_alpha}. As a consequence, the mode occupations~\eqref{diagonal_NESS} become independent of $s_\alpha$ and $r_\alpha$, being given simply by the average of the reservoir distribution functions:
\begin{equation}\label{mode_occ_symmetry}
	n(\omega_\alpha) = \frac{1}{2}\left [\bar{n}_c(\omega_\alpha) + \bar{n}_h(\omega_\alpha) \right ].
\end{equation}
This is again a function of the frequency $\omega_\alpha$ only, and therefore leads to stable current patterns as in the equilibrium case~\cite{Rivas2017}. Any distribution of defects that respects one of the above discrete symmetries will also yield $s_\alpha = r_\alpha$, leading to similar behaviour. 

Note that the relevant symmetries are determined both by the Hamiltonian and by the configuration of the baths driving the system out of equilibrium. For example, the transformation $\hat{\Theta}\hat{\Sigma}_x$ is a symmetry of the QWZ Hamiltonian, where $\hat{\Sigma}_x = \hat{R}_{\pi} \hat{\Sigma}_y$ is a reflection about the $x$ axis. However, because this reflection does not interchange the hot and cold baths it is irrelevant for the NESS. Therefore, defect distributions that are only symmetric under $\hat{\Sigma}_x$ destroy the bosonic edge currents whenever $T_c\neq T_h$.

Examples of the nonequilibrium distribution functions obtained for a symmetric defect configuration are shown in Fig.~\ref{fig:distributions}(a), corresponding to the system depicted in Fig.~4(b). As expected, the distributions are well behaved functions of frequency. Conversely, Fig.~\ref{fig:distributions}(b) shows that when the defect configuration does not respect either of the discrete symmetries, as in Figs.~4(c) or~\ref{fig:fermion_impurity_quiver}, the mode occupations are erratically varying functions of frequency. This disrupts the erasure effect and leads to the destruction of the boundary current patterns, in general. 

Fig.~\ref{fig:distributions} also shows that the fermionic distributions exhibit a sudden change at the chemical potential whenever $T_c$ is sufficiently small. This sharp feature explains the absence of counter-currents around the impurities in Figs.~4(b,c). In this case, the chemical potential is just below the particle-hole symmetric point $\mu=\omega_0$. This leads to a single ``hole'' that occupies an edge state, visible in Fig.~\ref{fig:distributions} as the isolated point just below $\omega_\alpha=\omega_0$. The edge current pattern observed in Fig.~4(b) is primarily due to this hole state which, occupying an edge mode, is robust against any bulk perturbation. Conversely, when the chemical potential is shifted down to $\mu = \omega_0- 0.5t$, several holes are created in other edge states [Fig.~\ref{fig:distributions}(b), blue circles], including edge states localized around the impurities. Consequently, we see the appearance of counter-currents shielding the impurities in Fig.~\ref{fig:fermion_impurity_quiver}(a). 

At higher temperature, the sharp feature in $n(\omega_\alpha)$ disappears [Fig.~\ref{fig:distributions}(b), light blue squares]. The current distribution is then determined mainly by bulk states and thus depends intimately on the symmetries of the impurity configuration. If the configuration is asymmetric, as in Fig.~\ref{fig:fermion_impurity_quiver}(b), the boundary currents are strongly disrupted. 

In the topologically trivial regime, there are no edge states and both bosons and fermions behave very similarly with respect to the introduction of defects. In particular, asymmetric impurity distributions tend to inhibit the formation of clean boundary current patterns. An example is shown in Fig.~\ref{fig:fermion_impurity_quiver}(c).

\begin{figure*}
	\includegraphics[width=0.99\linewidth]{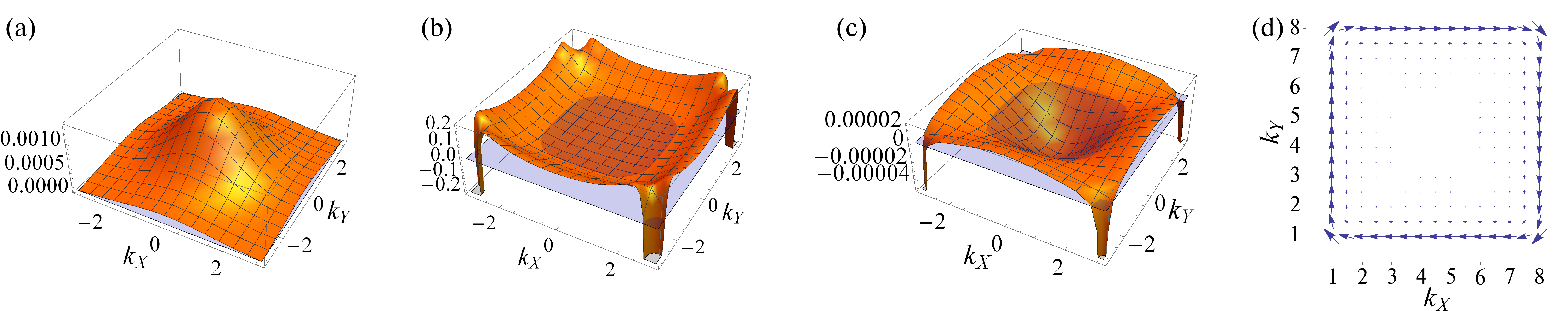}
	\caption{The emergence of edge currents in the weak coupling limit can be understood by a semiclassical argument. (a) The factor $n[\omega_-(\mathbf{k})]-n[\omega_+(\mathbf{k})]$ in Eq. \eqref{semi-classical_edge_current} for the bosonic case with $\omega_0=10t$, $T_h=t$, $T_c=0.01t$ and $m=2.1t$. (b) The Berry curvature ${\mathcal{F}\!\!}_-(\mathbf{k})$, which has a balanced positive and negative contributions over B.Z. (the zero value plane is depicted in light blue color) as for  $m>2t$ the model is in a trivial phase and ${\mathcal{F}\!\!}_-(\mathbf{k})$ integrates out to zero. (c) The product $\{n[\omega_-(\mathbf{k})]-n[\omega_+(\mathbf{k})]\}{\mathcal{F}\!\!}_-(\mathbf{k})$, which breaks such a balance, leads to a nonvanishing integral in Eq. \eqref{semi-classical_edge_current}. (d) The semiclassical current pattern generated according \eqref{semi-classical_edge_current} for the potential $V(x,y)=(x/8)^{10}+(y/8)^{10}$.  \label{fig:semiclassical}}
\end{figure*}

\subsection{Semiclassical picture}
\label{app:semiclassical}

The behaviour in the weak coupling can be understood by using a semiclassical picture \cite{Xiao2010,Price2012}.  Under periodic boundary conditions,  the QWZ Hamiltonian can be written in the form
\begin{equation}
\hat{H}=\sum_{\mathbf{k}\in \mathrm{B.Z.}} \mathbf{a}_{\mathbf{k}}^\dagger\cdot \mathbf{H}(\mathbf{k})\cdot \mathbf{a}_{\mathbf{k}},
\end{equation}
with $\mathbf{k}=(k_X,k_Y)$ the quasimomentum and
\begin{equation}
	\label{single-particle-Bloch-H}
\mathbf{H}(\mathbf{k})=\omega_0\mathbf{1}+\mathbf{h}(\mathbf{k})\cdot\bm{\sigma},
\end{equation}
where
\begin{equation}
\mathbf{h}(\mathbf{k})=[t_Y\sin(k_Y),t_X\sin(k_X),m+t_X\cos(k_X)+t_Y\cos(k_Y)].
\end{equation}
Note that we have taken a slightly different Hamiltonian with respect to the original reference \cite{Qi2006}.  The two energy bands of the model are then given by
\begin{equation}
\omega_{\pm}(\mathbf{k})=\omega_0\pm |\mathbf{h}(\mathbf{k})|,
\end{equation}
and for the isotropic case considered throughout this work  ($t_X=t_Y=t>0$),
\begin{equation}
|\mathbf{h}(\mathbf{k})|=t\sqrt{\sin^2(k_X)+\sin^2(k_Y)+[m/t+\cos(k_X)+\cos(k_Y)]^2}.
\end{equation}
The action of a force $\mathbf{F}$ modifies the semiclassical equation for the velocity $\mathbf{v}_\alpha$ of a Bloch wave-packet in the band $\omega_{\alpha}(\mathbf{k})$ (assuming the single-particle Hamiltonian~\eqref{single-particle-Bloch-H} is gapped,  i.e.  $m\neq0,\pm 2t$, and $m$ is small enough to justify the semiclassical approximation) as 
\begin{equation}\label{Vsem}
\mathbf{v}_\alpha=\frac{\partial \omega_\alpha(\mathbf{k})}{\partial\mathbf{k}}-\mathbf{F}\times \bm{\mathcal{F}}\!\!_\alpha(\mathbf{k}).
\end{equation}
Here, $\bm{\mathcal{F}}\!\!_\alpha(\mathbf{k})=[0,0,{\mathcal{F}\!\!}_\alpha(\mathbf{k})]$, with ${\mathcal{F}\!\!}_\alpha(\mathbf{k})$ the Berry curvature of the band, which, for a two-band model, can be written in the form
\begin{align}
{\mathcal{F}\!\!}_\pm(\mathbf{k})&=\mp\frac{1}{2}\mathbf{n}(\mathbf{k})\cdot\left[\frac{\partial\mathbf{n}(\mathbf{k})}{\partial k_X}\times \frac{\partial\mathbf{n}(\mathbf{k})}{\partial k_Y}\right],
\end{align}
with $\mathbf{n}(\mathbf{k}):=\mathbf{h}(\mathbf{k})/|\mathbf{h}(\mathbf{k})|$. For the QWZ Hamiltonian we find
\begin{equation}
{\mathcal{F}\!\!}_\pm(\mathbf{k})=\frac{\pm\big[\cos(k_X)+\cos(k_Y)+\tfrac{m}{t}\cos(k_X)\cos(k_Y)\big]}{2\big\{\!\sin^2(k_X)+\sin^2(k_Y)+\big[\tfrac{m}{t}+\cos(k_X)+\cos(k_Y)\big]^2\big\}^\frac32}.
\end{equation}
We can now argue for the presence of an edge circulation under open boundary conditions by introducing a confining potential $V(x,y)$ in the semiclassical equation for $\mathbf{v}_\pm$ \eqref{Vsem}.  Assuming that the potential varies slowly on the scale of the lattice, the energy of the wavepacket in the semi-classical approximation is simply $\omega_\alpha(\mathbf{k}) + V(x,y)$~\cite{Xiao2010}. Moreoever, the potential induces a confining force $\mathbf{F}=-\bm{\nabla}V(x,y)$,  which approximately vanishes in the bulk and becomes very large in the boundary of the confined region, 
\begin{align}
	\label{edge-bulk-velocity}
\mathbf{v}_\alpha^\mathrm{(bulk)}\simeq  \frac{\partial \omega_\alpha(\mathbf{k})}{\partial\mathbf{k}}, \quad \mathbf{v}_\alpha^\mathrm{(edge)}\simeq \bm{\nabla}V(x,y)\times \bm{\mathcal{F}}\!\!_\alpha(\mathbf{k}).
\end{align} 
Since $\mathbf{F}$ is normal and points inward from the confined boundary,  a circulating current is induced along this boundary with direction given the sign of the Berry curvature, which depends on the band $\alpha$, the sign of $m$, and the specific value of $\mathbf{k}$: clockwise for ${\mathcal{F}\!\!}_\alpha(\mathbf{k})>0$ and counterclockwise for ${\mathcal{F}\!\!}_\alpha(\mathbf{k})<0$. Thus, for the NESS at the weak coupling regime,  we may estimate the current density (current per unit width) by \cite{Xiao2010}
\begin{align}
	\label{semiclassical_current}
\mathbf{I}(x,y)=\sum_{\alpha=\pm}\int_{\mathrm{B.Z.}}d^2\mathbf{k}\, n[\omega_\alpha(\mathbf{k})+V(x,y)] \mathbf{v}_\alpha(x,y,\mathbf{k}),
\end{align}
where the integral extends over the first Brillouin zone. Here we have assumed that the NESS occupation number $n$ only depends on energy. As discussed in Sec.~\ref{app:distribution}, this is the case under a $\hat{\Pi}\hat{\Theta}\hat{\Sigma}_y$ or a $\hat{\Pi}\hat{R}_\pi$ symmetric configuration. 

Now, if we consider regions far from the boundary, so that $V(x,y)\ll \omega_\alpha(\mathbf{k})$, we may approximate Eq.~\eqref{semiclassical_current} by
\begin{align}
	\mathbf{I}(x,y)&\simeq\int_{\mathrm{B.Z.}}d^2\mathbf{k}\, \{n[\omega_-(\mathbf{k})]-n[\omega_+(\mathbf{k})]\} \mathbf{v}_-,
\end{align}
using the fact that $\mathbf{v}_+= -\mathbf{v}_-$. Since there is no privileged direction of $\mathbf{v}_-^{\mathrm{(bulk)}}$ on each energy shell: $\omega_\alpha(\mathbf{k})=\omega_\alpha(-\mathbf{k})$ and hence $\frac{\partial \omega_\alpha(\mathbf{k})}{\partial\mathbf{k}}=-\frac{\partial \omega_\alpha(-\mathbf{k})}{\partial\mathbf{k}}$. Therefore, the contribution of $\mathbf{v}_\alpha^{(\rm bulk)}$ to the integrand is odd and integrates to zero. We conclude that 
\begin{align}
	\label{semi-classical_edge_current}
	\mathbf{I}(x,y)&\simeq\int_{\mathrm{B.Z.}}d^2\mathbf{k}\, \left \{n[\omega_-(\mathbf{k})]-n[\omega_+(\mathbf{k})]\right \} \mathbf{v}_-^{(\rm edge)},
\end{align}
which clearly vanishes in the bulk of the system where $V(x,y)=0$ (see Eq.~\eqref{edge-bulk-velocity}).  Nevertheless, since $n(\omega)$ is a monotonically decreasing function of energy [Eq.~\eqref{mode_occ_symmetry}], there is an imbalance of the contributions to $\mathbf{I}(x,y)$ within the Brillouin zone in regions where $\mathbf{v}_\alpha^{\rm edge} \propto \nabla V(x,y) \neq 0$. Thus, Eq.~\eqref{semi-classical_edge_current} predicts a net chiral edge current,  see Fig. \ref{fig:semiclassical}. This erasure effect sustains the appearance of the chiral current in the symmetric situation, but it is spoilt in the nonsymmetric case,  as the occupation number is no longer only a function of the energy alone.

%
%
%\bibliographystyle{apsrev4-2}
%\makeatletter
%\renewcommand\@biblabel[1]{[S#1]}
%\makeatother
%\bibliography{bibliography}
%\clearpage
%
%\end{document}

\end{document}